\documentclass{IEEEtaes}
\usepackage{cite}
\usepackage{amssymb}
\usepackage{amsmath}
\usepackage{amsfonts}
\usepackage{array}
\usepackage{multicol}
\usepackage{textcomp, gensymb}
\usepackage{mathtools}
\usepackage{graphicx}
\usepackage{subfigure}
\usepackage{bm}
\usepackage{comment}
\usepackage{color}
\usepackage{xcolor}
\usepackage{multirow}
\usepackage{mathtools,amssymb,lipsum}

\usepackage{cuted}
\usepackage[nolist,nohyperlinks]{acronym}

\newcommand{\bzeta}{\bm{\zeta}}

\newcommand{\btheta}{\bm{\theta}}
\newcommand{\balpha}{\bm{\alpha}}
\newcommand{\bgamma}{\bm{\gamma}}
\newcommand{\bkappa}{\bm{\kappa}}
\newcommand{\bomega}{\bm{\omega}}
\newcommand{\bOmega}{\bm{\Omega}}

\newcommand{\E}{\mathbb{E}}

\DeclareMathOperator*{\argmax}{arg\,max}
\DeclareMathOperator*{\argmin}{arg\,min}

\usepackage{color}  
\def\cred{\textcolor{black}}

\jvol{XX}
\jnum{XX}
\jmonth{XXXXX}
\paper{1234567}
\pubyear{2020}
\doiinfo{TAES.2020.Doi Number}

\newcommand{\haoqing}[1]{\textcolor{red}{\textbf{[{\em Haoqing SAYS}: #1]}}}

\begin{document}

\title{Robust Interference Mitigation techniques for Direct Position Estimation} 


\author{Haoqing Li}
\member{}
\affil{Northeastern University, Boston, MA 02115, USA} 
\author{Shuo Tang}
\member{}
\affil{Northeastern University, Boston, MA 02115, USA} 
\author{Peng Wu}
\member{}
\affil{Northeastern University, Boston, MA 02115, USA} 

\author{Pau Closas}
\member{Senior Member, IEEE}
\affil{Northeastern University, Boston, MA 02115, USA} 


\receiveddate{
This work has been partially supported by the National Science Foundation under Award ECCS-1845833.}


\authoraddress{Authors are with the Department of Electrical and Computer Engineering at Northeastern University, 360 Huntington Avenue, Boston, MA 02115 (USA). e-mail: \{li.haoq, tang.shu,  wu.p,  closas\}@northeastern.edu}


\markboth{LI ET AL.}{ROBUST INTERFERENCE MITIGATION TECHNIQUES FOR DIRECT POSITION ESTIMATION}
\maketitle

\acrodef{GPS} {Global Positioning System}
\acrodef{GNSS} {Global Navigation Satellite System}
\acrodef{PVT}{Position Velocity and Time}
\acrodef{CAF}{Cross Ambiguity Function}
\acrodef{DLL}{Delay Lock Loop}
\acrodef{PLL}{Phase Lock Loop}
\acrodef{PAC}{Pulse Aperture Correlator}
\acrodef{NN}{Neural Network}
\acrodef{NNs}{Neural Networks}
\acrodef{DNN}{Deep Neural Network}
\acrodef{MLP}{Multilayer Perceptron}
\acrodef{PRN}{Pseudo-Random Noise}
\acrodef{ROC}{Receiver Operating Characteristic}
\acrodef{CN0}{carrier-to-noise-density ratio}
\acrodef{RF}{Radio Frequency}
\acrodef{AWGN}{Additive White Gaussian Noise}
\acrodef{iid}[i.i.d.]{independent and identically distributed}
\acrodef{PSD}{Power Spectral Density}
\acrodef{SMR}{Signal-to-Multipath Ratio}
\acrodef{LOS} {Line-of-Sight}
\acrodef{NLOS} {None-Line-of-Sight}
\acrodef{E-L}{Early-Late}
\acrodef{HRC} {High Resolution Correlatior}
\acrodef{CIR} {Chanel Impulse Response}
\acrodef{CDF} {Cumulative Density Function}
\acrodef{ANFIS}{Adaptive Neuro Fuzzy Inference System}
\acrodef{CNN}{Convolution Neural Network}
\acrodef{SVM}{Support Vector Machine}
\acrodef{ANN}{Artificial Neural Network}
\acrodef{GBDT}{Robust Gradient-Boosting Decision Tree}
\acrodef{DT}{Decision Tree}
\acrodef{SVM}{Support Vector Machine}
\acrodef{GPR}{Gaussian Process Regression}
\acrodef{LSTM}{Long Short-Term Memory}
\acrodef{C-SVM}{classification SVM}
\acrodef{PCA}{Principal Component Analysis}
\acrodef{BPNN}{Back-Propagation Neural Network}
\acrodef{RNN}{Recursive Neural Network}
\acrodef{AI} {Artificial Intelligence}
\acrodef{SINS}{Strapdown Inertial
Navigation System}
\acrodef{DPE}{Direct Position Estimation}
\acrodef{ZZB}{Ziv-Zakai Bound}
\acrodef{RIM}{Robust Interference Mitigation}
\acrodef{2SP}{two-step positioning}
\acrodef{IC}{Interference Cancellation}
\acrodef{PDF}{Probability Density Function}
\acrodef{TD}{Transformed Domain}
\acrodef{FD}{Frequency Domain}
\acrodef{ZMNL}{Zero Memory Non-Linearity}
\acrodef{ML} {Maximum Likelihood}
\acrodef{DME}{Distance Measuring Equipment}
\acrodef{DD-RIM}{Dual Domain RIM}
\acrodef{CRB}{Cramér–Rao Bound}
\acrodef{CW}{Continuous Wave}
\acrodef{ARNS}{Aeronautical Radionavigation Services}
\acrodef{ppps}{pulse pairs per second }
\acrodef{SNR}{signal-to-noise-ratio}
\acrodef{FIM}{Fisher Information Matrix}
\acrodef{IFIM}{Instantaneous Fisher Information Matrix }
\acrodef{CLT}{Central Limit Theorem}
\acrodef{MQBD}{mean quadratic bandwidth}
\acrodef{FFT}{Fast Fourier Transform}
\acrodef{IFFT}{Inverse Fast Fourier Transform}
\acrodef{DD}{Dual Domain}
\acrodef{LoE}{Loss of Efficiency}
\acrodef{LS}{Least Square}
\acrodef{RMSE}{Root Mean Square Error}
\acrodef{FSPLM}{ Free-Space Path Loss Model}
\acrodef{ARS}{Accelerated Random Search}
\acrodef{SD}{Single Domain}
\acrodef{SD-RIM}{Single Domain RIM}
\acrodef{JN}{Jamming to Noise ratio}
\acrodef{MAR}{Median Absolute Deviation}


%

\begin{abstract}
\ac{GNSS} is pervasive in navigation and positioning applications, where precise position and time referencing estimations are required. Conventional methods for \ac{GNSS} positioning involves a two-step process, where intermediate measurements such as Doppler shift and time delay of received \ac{GNSS} signals are computed and then used to solve for the receiver's position. Alternatively, \ac{DPE} was proposed to infer the position directly from the sampled signal without intermediate variables, yielding to superior levels of sensitivity and operation under challenging environments. However, the positioning \cred{resilience} of \ac{DPE} method is still under the threat of various interferences. \ac{RIM} processing has been \cred{studied} and proved to be efficient against various interference in conventional \ac{2SP} methods, and therefore worthy to be explored \cred{regarding} its potential to enhance \ac{DPE}. This article extends \ac{DPE} methodology by incorporating \ac{RIM} strategies that address the increasing need to protect \ac{GNSS} receivers against intentional or unintentional interferences, such as jamming signals, which can deny \ac{GNSS}-based positioning. \ac{RIM}, which leverages robust statistics, was shown to provide competitive results in \cred{two-step} approaches and is here employed in a high-sensitivity \ac{DPE} framework with successful results.
The article also provides a quantification of the loss of efficiency of using \ac{RIM} when no interference is present and validates the proposed methodology on relevant interference cases, \cred{while the approach can be used to mitigate other common interference signals}.  
\end{abstract}


\section{Introduction}
The conventional approach to process \ac{GNSS} signals is a \ac{2SP} process, 
where the so-called \ac{CAF} is computed and maximized as a function of time delay and Doppler shift of each in-view satellite \cite{kaplan2017understanding, borre2005software,morton2021position}. The \ac{GNSS} solution including position and velocity of a \ac{GNSS} receiver is then calculated based on the time delay and Doppler shift from the first step. Despite of the generality and efficiency of the \ac{2SP} approach, the fact that intermediate measurements (Doppler shift and time delay) are used would degrade the performance compared with the case when position is directly estimated in one step, which is the \ac{DPE} approach.  This is proved in \cite{iltis1994adaptive, amar2008new} showing the performance of \ac{DPE} approach can never be worse than the \ac{2SP} approach. One of the main benefits of DPE processing is that receivers can increase their senstivity, thus being able to operate at lower signal-to-noise ratios compared to their 2SP versions \cite{closas2021direct}.

\ac{DPE} for \ac{GNSS} was first proposed in \cite{closas2007maximum}. This approach is based on the fact that time delays and Doppler
shifts of all satellites are intimately related to one another through the \ac{GNSS} solution of receiver. Considering the \ac{CAF} as a function of \ac{PVT} of \ac{GNSS} receiver, the \ac{PVT} results can be acquired in just one step by maximizing the \ac{CAF}. 
Compared with conventional \ac{2SP} approach, \ac{DPE} approach has following advantages \cite{closas2021direct}: $i)$ no intermediate measurements: as discussed above, \ac{2SP} approach needs to estimate Doppler shift and time delay parameters for \ac{GNSS} solution, which brings potential correlation among channels and propagation effects. Those errors would cause further distortions in \ac{GNSS} solution through non-linearity; $ii)$ lower dimension size: since \ac{2SP} approach needs to estimate Doppler shift and time delay of every available channel, the dimension size can simply increase to a larger value especially in a multi-constellation receiver, while \ac{DPE} approach only need to estimate \ac{PVT} solution; $iii)$ simpler syncronization problem: The prior information from the tracking loops in \ac{2SP} approach is generally applied as an involved task \cite{pany2010navigation}
and would need extensive test-field campaigns to generate relevant data \cite{steingass2004measuring}, and the algorithm needs to clarify among difference synchronization evolution models based on the dynamics of receiver. This is much more difficult compared with the case when the parameter of interest is the user’s position itself, with each parameter has their own  physical meaning to aid the inclusion of side information; $iv)$ robustness, \ac{DPE} approach is more robust than \ac{2SP} approach against interferences, given the estimation of position is jointly performed taking into account measurements from all in–view satellites \cite{closas2007maximum, closas2015evaluation}.

However, regardless of the robustness of \ac{DPE} compared with \ac{2SP} approach, the interferences can still cause a degradation to its performance. Such interferences, such as intentional jammers or unintentional interferences \cite{amin2016vulnerabilities}, become challenging threats in the \ac{GNSS} processing chain. Despite of the fact that jammers are illegal devices in most (not all) countries, they are very easy to built and cheap to buy, those devices can cause a large-area disruption to GNSS-based services (in kilometers level). In addition, unintentional interferences can also be a problem in GNSS positioning. For example, the \ac{DME} signal, which is essential in aircraft navigation, or other technologies are known to interfere GNSS signals \cite{gao2007dme,ioannides2016known,arribas2019air}. Therefore, the research of interference mitigation techniques have been triggered recently.

In terms of the \ac{2SP} approach, a classical jamming signal mitigation method is \ac{IC}, in which the detection, estimation and reconstruction of the interference waveform is done. For instance, pulse blanking and (adaptive) notch filtering \cite{borio2013optimal, borio2008two} are the two typical and popular \ac{IC} methods. However, the drawback of this approach cannot be neglected, where detection and estimation are two
possible causes of failure during processing, and that there is a need to make assumptions on the jamming signal waveform \cite{boriognss}.
To overcome those drawbacks, a robust statistics based approach was investigated, where interferences are regarded as outliers. It is referred to as \ac{RIM} approach, in which  the estimation of the interference waveform and its detection can be avoided. The concept was first implemented  in \cite{borio2017myriad}, where the RIM approach acts as a filter to mitigate pulsed interference as outliers in received signal. In \cite{borio2017myriad}, the myriad \ac{ZMNL} was derived by substituting the classical Gaussian assumption with Cauchy assumption  on the jammed input signal, while the complex signum \ac{ZMNL}, is derived under a Laplacian model in \cite{daniele2018complex}. Both works apply their \ac{ZMNL}s in time domain, under a more relaxed assumption of heavy-tailed \ac{PDF} to the noise statistics, modelling large outliers in the sampled signal. Then, \cite{borio2018huber, borio2018huberm, li2018robust} explored the use of Huber’s \ac{ZMNL} in transformed domain instead of the time one. Furthermore, \cite{li2019dual} studied Huber’s \ac{ZMNL} in multiple domains, both time and transformed, which was referred to as Dual-Domain RIM. Recently, \cite{gioia2021multi} has discussed the jointly use of \ac{RIM} approach and other typical interference mitigation techniques in multi-layer multi-constellation \ac{GNSS} processing.
In this paper, we study the potential of \ac{RIM} approach in \ac{DPE} processing, considering Huber's \ac{ZMNL} to single- and dual-domains, exploring the performances of them in the presence and absence of different kinds of jamming signals. Specifically, intentional \ac{CW} jamming signal and \ac{DME} interference signal. 
\cred{Notice that RIM, which DPE-RIM is based on, is effective against interference signals that can be considered to be outliers in time (e.g. pulsed interferences), frequency (e.g. CW), in arbitrary domains (e.g. wavelet transform), or in multiple domains (e.g. the case of the DME signal) \cite{li2019dual}.
As a consequence, RIM (and therefore DPE-RIM) could be applicable to wideband interferences when these appear as pulsed in time domain, otherwise this methodology is not applicable and other solutions such as the use of antenna arrays may be considered \cite{fernandez2016robust}.}

In summary, the main novel contributions of this article with respect to previously published works are:
\begin{itemize}
    \item A robust \ac{DPE} receiver solution that mitigates interferences through the incorporation of \ac{RIM} methodology. This results in a novel GNSS receiver framework that features high-sensitivity and interference rejection.
    \item Analysis of the \ac{LoE} of such approach in terms of the \ac{CRB} degradation under a direct-positioning framework.
    \item Performance analysis of a specific \ac{RIM} method (based on Huber's non-linearity) under direct-positioning framework against \ac{CW} and \ac{DME} interferences, validating the theoretical results.
\end{itemize}

The remainder of the paper is organized as follows: Section~\ref{Sec:Backgrounds} describes the signal model for both \ac{GNSS} and interference signals, as well as recalls the basics of \ac{DPE} processing. Section~\ref{Sec:RIM} contains the main contribution, showing the application of \ac{RIM} to \ac{DPE}.  Section~\ref{Sec:LoE} provides a discussion and the derivation of \ac{LoE} of \ac{RIM} under \ac{DPE}, which corresponds to the degradation of using \ac{RIM} when there is no interference present. Section~\ref{Sec:SimSetup} details the simulation experiments 
and corresponding analysis. Finally, the paper concludes with final remarks in Section~\ref{Sec:Conclusions}.

\section{Signal models and direct-positioning background}\label{Sec:Backgrounds}

This section provides a discussion on the signal models for GNSS signals and interference signals that are used later in the article. Particularly, we formulate the \cred{signal} model for a generic intentional jammer and the DME signal, the latter being a type of unintentional interference that is explored in the simulations result. This section also provides a review of \ac{DPE} signal processing, which will be augmented with the \ac{RIM} approach in Section \ref{Sec:RIM}. 

\subsection{Signal Model}
As described in \cite{closas2017direct, closas2021direct}, the complex baseband equivalent of the received signal at an antenna can be modeled as the summation of several scaled, structure-known signals with time delay and Doppler shift as shown below:
\begin{equation}\label{eq:2spsignalmodel}
    x(t) = \sum\limits_{i=1}^M \alpha_i c_i(t-\tau_i)\mbox{e}^{j(2\pi f_{d,i}t + \phi_i)} + \eta(t) + i(t)
\end{equation}
where $M$ is the number of satellites that are visible to the receiver, the index $i\in \{1, \cdots\, M\}$ denotes each satellite, $\alpha_i$ is the complex amplitude containing phase information, $c_i(t)$ is the complex navigation signal spread by the corresponding \ac{PRN} code, $\tau_i$ is the time delay from the satellite to the receiver, $f_{d,i}$ is the Doppler shift, $\eta(t)$ denotes \ac{AWGN} signal with double sided spectral density $N_0/2$, and  $\phi_{i}$ denotes the phase shift introduced by the communication channel, which is regarded as an unknown parameters alongside $\tau_i$ and $f_{d,i}$. In the absence of interference $\eta(t)$ is the dominating random term and the reason for assuming that $x(t)$ follows a Gaussian distribution. When an interference is present, $i(t)$, the noise component $\eta(t) + i(t)$ would incorporate both contributions. In this paper, $i(t)$ is modeled as a generic signal, and can be, for instance, a \ac{CW} jamming signal or a \ac{DME} interference signal as will be described later in this section. The covariance of the noise is defined as
\begin{equation}
    \E\{\eta(t)\eta(t)^H\} = \sigma_n^2 \;.
\end{equation}
It is noted that the noise signal is circularly-symmetric complex Gaussian such that the real and imaginary parts have the same variance $\sigma_n^2/2$. After sampling at a suitable rate $f_s = \frac{1}{T_s}$ that satisfies the Nyquist criterion, the resulting complex discrete-time sequence is: 
\begin{equation}
	x[n] = \sum\limits_{i=1}^M \alpha_i c_i(nT_s-\tau_i)\mbox{e}^{j(2\pi f_{d,i}nT_s + \phi_i)} + \eta[n] + i[n]
	\label{signalmodel}
\end{equation}

\subsection{Interference signal}
\label{ssec.interfsig}
The baseband interference signal, $i[n]$, can assume different forms depending on the type of source generating it \cite{borio2016impact,morales2019survey}. 
A wide class of interference signals can be modeled as
\begin{equation}
i[n] = \alpha_I \mbox{e}^{j2\pi f_I[n]nT_s + j\phi_I[n]} \;,
\label{eq:campSignals}
\end{equation}
that is, signals with a constant amplitude, $\alpha_I$, and a time-varying frequency/phase, $f_I[n]$ and $\phi_I[n]$ respectively. For instance, \ac{CW} interferences can be modeled as \eqref{eq:campSignals} with some constant parameters $f_I[n] = f_{CW} = \mbox{const}$ and $\phi_I[n] = \phi_{CW} = \mbox{const}$.
%
%
When the interference amplitude is assumed constant, the signal model \eqref{eq:campSignals} is not able to capture pulsed signals such as DME components. Due to its relevance in the context of \ac{GNSS}, \ac{DME} signals are specifically discussed in the next section.

\subsection{Distance Measurement Equipment signal model}
\ac{DME} is used to measure the distance between aircraft and ground station by measuring the propagation delay between a \ac{DME} interrogator equipment onboard the aircraft and a transponder at the ground station. It operates in four modes: X, Y, W and Z, between $960$ MHz and $1215$ MHz in an \ac{ARNS} band \cite{gao2007dme}. In particular, the X-mode replies in $1151-1213$ MHz, which thus overlaps with the \ac{GNSS} $E5$ and $L5$ bands. For this reason, \ac{GNSS} signal reception in the $E5$ and $L5$ band can be degraded by \ac{DME} signals. Therefore, \acp{DME} replying in X-mode can interfere with \ac{GNSS} signal reception and should be mitigated. 

\begin{figure}
\centering
  \includegraphics[width=1.00\columnwidth]{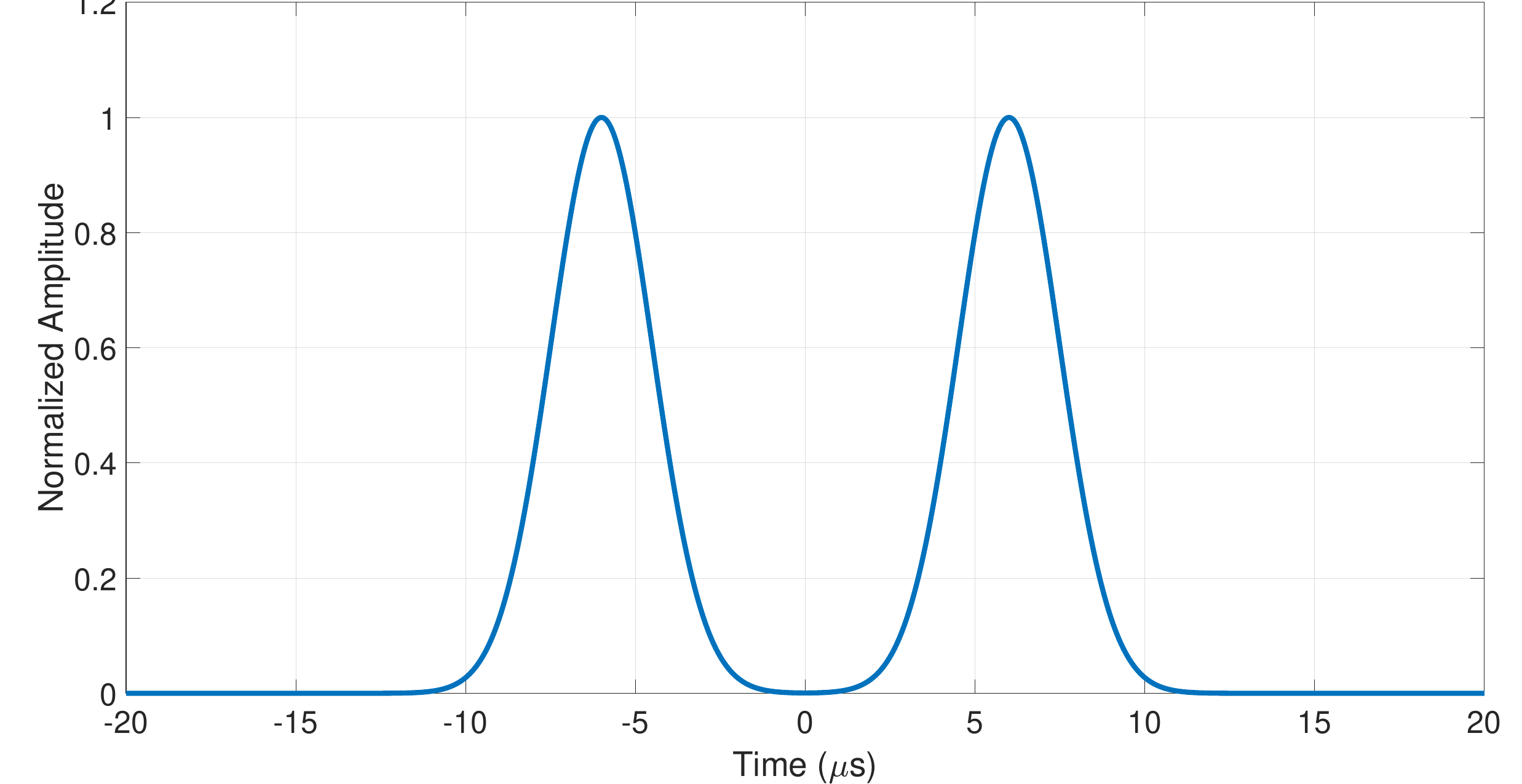}
  \caption{Example of \ac{DME} signal waveform with normalized amplitude.}
	\label{fig::DME_Simulation}
\end{figure}

\ac{DME} signals are composed of pulse pairs and Fig.~\ref{fig::DME_Simulation} shows one pair of \ac{DME} signal in time domain. As shown in the figure, considering its short time duration and high peak power, the \ac{DME} signal can be regarded as an outlier in the time domain. Moreover, when considering its \ac{PSD} in \cite{gao2007dme}, \ac{DME} signals can also be considered as an outlier in the frequency domain due to its high power concentrated in a narrow band. More details of the parameters and modeling of \ac{DME} signals can be found in \cite{gao2007dme, epple2012modeling}.

\subsection{Direct Position Estimation}
The signal model in \eqref{eq:2spsignalmodel}, typically considered in most receiver designs \cite{morton2021position}, assumed that delay and Doppler are constant within an observation window (the integration interval). However, in practice, these quantities evolve over time as a consequence of their physical interpretation \cite{closas2009bayesian}. We review this in this section, while we notice that time delay and Doppler shifts can be parameterized by the position of the receiver, as well as the time-varying positions and velocities of the satellites. Particularly, the time delay -- or the signal propagation time --, is related to the distance between the satellite and the receiver. Consequently, the pseudorange observable $\rho_i = c\tau_i$ is modeled as
 \begin{equation}\label{eq:pseudorange}
     \rho_i = \varrho_i(\bm{p}) + c(\delta t-\delta t_i) + \epsilon_i,
 \end{equation}
 \begin{center}
 \begin{tabular}{ p{0.8cm}p{6.8cm}}
  $\varrho_i(\bm{p})$ & $||\bm{p}-\bm{p}_i||$ between the $i$-th satellite, located at position $\bm{p}_i = (x_i, y_i, z_i)^\top$, and the receiver, whose position $\bm{p} = (x, y, z)^\top$ is unknown; \\
  $c$ & is the speed of light in m/s; \\
  $\delta t$ & the unknown receiver clock bias with respect to GNSS time; \\
  $\delta t_i$ & the $i$-th satellite clock bias with respect to GNSS time given by the ephemeries; and \\
  $\epsilon_i$ & a random term including ephemeris errors, atmospheric-induced delays, relativistic effects, and other unmodeled errors.
 \end{tabular}
 \end{center}

 The Doppler shift is the difference between the observed carrier frequency and its nominal value at transmission. The Doppler effect is caused by the relative motion between the receiver and the corresponding satellite. The Doppler shift can be modeled as
 \begin{equation}\label{eq:doppler}
     f_{d,i} = -(\bm{v}_i-\bm{v})^\top\bm{u}_i(1+\dot{\delta t})\frac{f_c}{c},
 \end{equation}
where $\bm{v}_i = (v_{x,i}, v_{y,i}, v_{z,i})^\top$ is the velocity vector of the $i$-th satellite, $\bm{v} = (v_x, v_y, v_z)^\top$ is the velocity of the receiver, $\dot{\delta t}$ is the clock drift of the receiver, and $\bm{u}_i$ denotes the unit vector from the receiver pointing to the $i$-th satellite as 
     $\bm{u}_i = \frac{\bm{p}_i - \bm{p}}{||\bm{p}_i - \bm{p}||}$ ,
 where $||\cdot||$ denotes the $\ell_2$-norm of a vector and $f_c$ denotes the carrier frequency of the transmitted GNSS signal.

As shown in \eqref{eq:pseudorange} and \eqref{eq:doppler}, the delay $\tau_i$ and Doppler shift $f_{d,i}$ of the $i$-th satellite are functions of the position $\bm{p}$ and velocity $\bm{v}$ of the receiver. More generally, if we gather all dynamics-related unknown parameters into a vector $\bkappa$ (for instance, $\bkappa = \bm{p}$ or $\bkappa^\top = ( \bm{p}^\top, \bm{v}^\top )$ \cite{closas2021direct}), the signal model in \eqref{eq:2spsignalmodel} can be parameterized by $\bkappa$ 
\begin{equation}\label{eq:dpesignalmodel}
    x(t) = \sum\limits_{i=1}^M \alpha_i c_i(t-\tau_i(\bkappa))\mbox{e}^{j(2\pi f_{d,i}(\bkappa)t + \phi_i)} + \eta(t) + i(t)
\end{equation}
After sampling at a $f_s = \frac{1}{T_s}$ that satisfies the Nyquist criterion, the resulting discrete-time complex signal is: 
\begin{equation}
	x[n] = \sum\limits_{i=1}^M \alpha_i c_i(nT_s-\tau_i(\bkappa))\mbox{e}^{j(2\pi f_{d,i}(\bkappa)nT_s + \phi_i)} + \eta[n] + i[n] 
\label{eq:simpleModelDigi}
\end{equation}

\ac{DPE} solves for the \ac{ML} estimation of $\bkappa$, given the model in \eqref{eq:dpesignalmodel}. It can be seen \cite{closas2021direct} that maximizing such likelihood is equivalent to minimizing the cost function:
\begin{equation}
    \Lambda(\bm{\kappa}) =\sum_{n=0}^{N-1}|x[n] - \sum_{i=1}^{M}\alpha_ic_i(nT_{s}-\tau_{i}(\bkappa))\mbox{e}^{j(2\pi f_{d,i}(\bkappa)nT_{s}+\phi_i)}|^2
    \label{eq:CF}
\end{equation}
Following the derivation from \cite{closas2021direct}, the estimate of $\bkappa$ is
\begin{equation}\label{eq:MLEabs}
\small
\begin{aligned}
\hat{\bkappa} 
    & = \argmax_{\bkappa}\left\{\sum\limits_{i=1}^M \left|\sum_{n=0}^{N-1}x[n]c_i(nT_{s}-\tau_{i}(\bkappa))\mbox{e}^{-j(2\pi f_{d,i}(\bkappa)nT_{s})}\right|^2\right\}\\
    &= \argmax_{\bkappa}\left\{\sum\limits_{i=1}^M \left|\mathcal{C}_i(\bkappa)\right|^2\right\}
\end{aligned}
\end{equation}
where $\mathcal{C}_i(\bkappa)$ is the so-called \ac{CAF} of the $i$-th satellite \cite{morton2021position}, defined as the correlation between the received samples and the local code, in this case parameterized by $\bkappa$.
Notice that when carrier-phase is also parameterized by $\bkappa$ the resulting cost function would be different, in which case high-accuracy DPE would be enabled \cite{TangPLANS23}. In this work we restrict to the typical DPE case where phase is considered unknown but independent of $\bkappa$ \cite{closas2021direct}.

\section{Robust Interference Mitigation and direct-positioning}
\label{Sec:RIM}
This section discusses how the \ac{RIM} approach can be incorporated in \ac{DPE} approach. Particularly, its applicability in various domains is treated, namely: \ac{TD} and \ac{DD}. 
At a glance, \ac{RIM} modifies the maximum likelihood cost function that typically results in \eqref{eq:MLEabs} using a nonlinear function $\rho(\cdot)$, which produces estimates that are more robust to outliers. In this case, outliers are interference signals that are stronger than \ac{GNSS} signals and sparse in one or several of the aforementioned domains \cite{boriognss}. In practice, RIM results in a variation of the CAF, which is referred to as a Robust \ac{CAF} and denoted as $\mathcal{C}_{\rho, i}(\bkappa)$ for the $i$-th satellite (cf. Appendix~\ref{sec:RIMDPE}). In the context of \ac{DPE}, the resulting robust estimation of the parameters in $\bkappa$ is then
\begin{equation}\label{eq:robustMLEabs}
\hat{\bkappa} 
    = \argmax_{\bkappa}\left\{\sum\limits_{i=1}^M \left|\mathcal{C}_{\rho, i}(\bkappa)\right|^2\right\}
\end{equation}
\noindent where the definition of depends on the time of \ac{RIM} processing performed, as detailed in the following subsections. Notice that, in the case of \ac{2SP}, the \ac{RIM} solution resembles \eqref{eq:robustMLEabs} with the exception that there is no sum over satellites and that the CAF is parameterized by time delay and Doppler shift instead of $\bkappa$.

\subsection{RIM in TD} \label{sec:TDRIM}
\begin{figure}
	\centering
		\includegraphics[width=0.8\columnwidth]{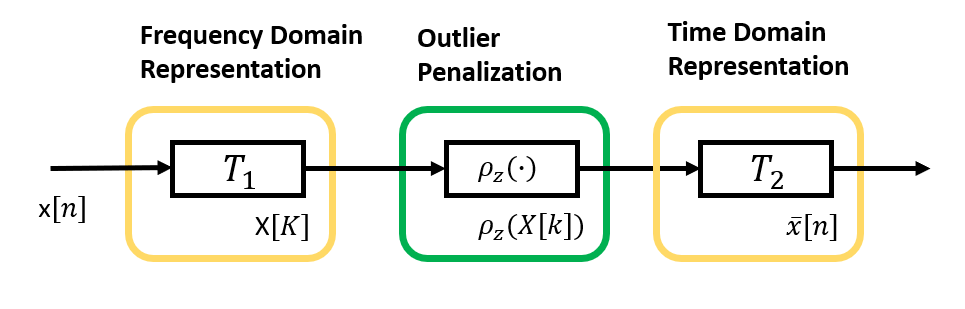}
	\caption{Generic block diagram of \ac{RIM} processing on signal samples.}
	\label{fig:genproscheme}
\end{figure}

In \ac{RIM} processing, the \ac{ZMNL}s can be applied in general \acp{TD}, which is depicted in Fig.~\ref{fig:genproscheme}.
A linear transform, $\mathbf{T}_1$, is used to project the interference component into a domain such that it occurs as a sparse representation, where only a limited number of samples are affected.
Transform $\mathbf{T}_1$ produces the \ac{TD} samples
\begin{equation}
		X[k] = \mathbf{T}_1(x[n]).
\label{eq:tdsamples}
\end{equation}
The change of index, from $n$ to $k$, is a notational convention adopted to indicate that the input samples, $x[n]$, have been brought to a different representation domain. 
%
Following $\mathbf{T}_1$, a \ac{ZMNL} is used to reduce the impact of outliers in the \ac{TD}.
A generic \ac{ZMNL} is denoted here as ${\rho_z}(\cdot)$ and produces the samples
\begin{equation}
	X_{\rho_z}[k] = {\rho_z}(X[k]) \;.
\label{eq:nlsamples}
\end{equation}
Finally, a second linear transform, $\mathbf{T}_2$, is applied to the samples, $X_{\rho_z}[k]$ to obtain new, filtered time domain discrete-time signal. $\mathbf{T}_2$ inverts the effects of $\mathbf{T}_1$ and brings back the samples to the time domain.
The output of $\mathbf{T}_2$ is denoted here as
\begin{equation}
\bar{x}[n] = \mathbf{T}_2(X_{\rho_z}[k]).
\label{eq:finalsample}
\end{equation}
Therefore we can say that $\mathbf{T}_1$ and $\mathbf{T}_2$ are inverse operators, $\mathbf{T}_1 \circ \mathbf{T}_2 = \mathbf{I}$, 
where $\mathbf{I}$ is the identity operator.
Note that the above \ac{TD} formulation is general and encompasses different alternatives such as time domain (when both $\mathbf{T}_1$ and $\mathbf{T}_2$ are identity operators) or frequency domain (when $\mathbf{T}_1$ is a \ac{FFT} matrix and $\mathbf{T}_2$ is the \ac{IFFT}) processing. \ac{RIM} aims at reducing the impact of an interference $i[n]$ on the cleaned samples, $\bar{x}[n]$, which are used for the computation of the robust \ac{CAF} \cite{borio2017myriad}. Following the procedure shown in Appendix~\ref{sec:RIMDPE}, a robust \ac{CAF} after TD-RIM can be computed as: 
\begin{equation}
	\mathcal{C}_{\rho,i}(\bkappa) = \sum_{n = 0}^{N - 1}\bar{x}[n]c_i\left(nT_s - \tau_i(\bkappa)\right)\mbox{e}^{-j2\pi f_{d,i}(\bkappa) nT_s},
\label{eq:robcaf}
\end{equation}
\noindent which can be then used to solve for \ac{DPE}'s positioning solution in \eqref{eq:robustMLEabs}.

In essence, the robust CAF $\mathcal{C}_{\rho, i}(\bkappa)$ 
applies a preprocessing to the data by means of a nonlinear function $\rho_z(x[n])$.
A variety of nonlinearities can be employed to pre-process the signal that constructs the so-called robust \ac{CAF}, as reviewed in Appendix \ref{sec:RIMZMNL}. 

%
\subsection{RIM in DD}
\begin{figure}
	\centering
		\includegraphics[width=0.95\columnwidth]{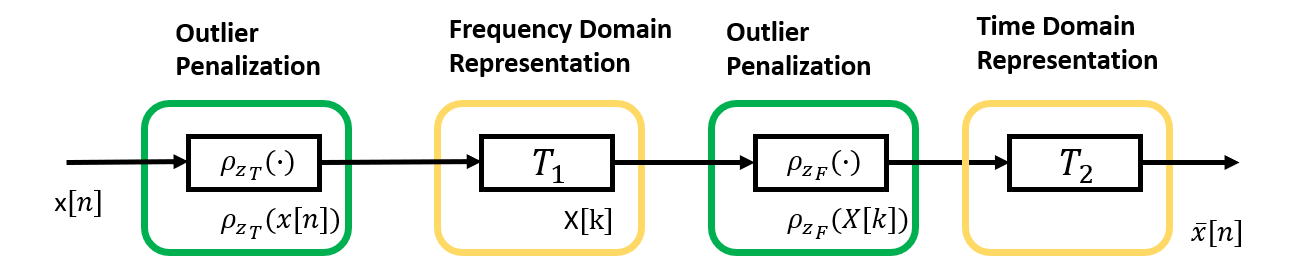}
	\caption{\acs{DD-RIM} with \ac{ZMNL}s applied successively in the time and frequency domains. Green boxes indicate the application of a \ac{ZMNL} and yellow boxes denote a linear transformation of the signal.}
	\label{fig:Huber_DD_TF}
\end{figure}

Following the approach proposed in \cite{li2019dual} for two-steps processing schemes, this section describes the implementation of \ac{ZMNL}s in two consecutive domains, referred to as \acs{DD-RIM}. It can be regarded as a cascade of two \ac{TD} \ac{RIM} processing blocks, for instance time and frequency domains. In particular, a doubly robust \ac{CAF} is obtained as follows:
\begin{equation} 
\mathcal{C}_{\rho,i}(\bkappa)= \sum_{n=0}^{N-1}\bar{x}[n] c_i\left(nT_s - \tau_i(\bkappa)\right)\mbox{e}^{-j2\pi f_{d,i}(\bkappa) nT_s}.
\label{DD_CAF}
\end{equation}
where $\bar{x}[n]$ are the time domain samples obtained after the sequential nonlinear processing on time and transformed domains as shown in Fig.~\ref{fig:Huber_DD_TF}, mathematically described as
\begin{equation}
\bar{x}[n] = \mathbf{T}_2\left({\rho_z}_{F}\left(X[k]\right)\right)
\label{eq:freq2time}
\end{equation}
where 
\begin{equation}
X[k] = \mathbf{T}_1\left({\rho_z}_{T}\left(x[n]\right)\right)
\label{eq:time2freq}
\end{equation}
In this section, $\mathbf{T}_1$ and $\mathbf{T}_2$ are specified as \ac{FFT} and \ac{IFFT}, to bring the signal from the time to the frequency domains and vice versa. In general, other pairs of transformations could be used \cite{boriognss}.
Estimates of the signal parameters are then obtained by maximizing the robust \ac{CAF} as in \eqref{eq:MLEabs}.

\begin{figure}
	\centering
		\includegraphics[width=0.95\columnwidth]{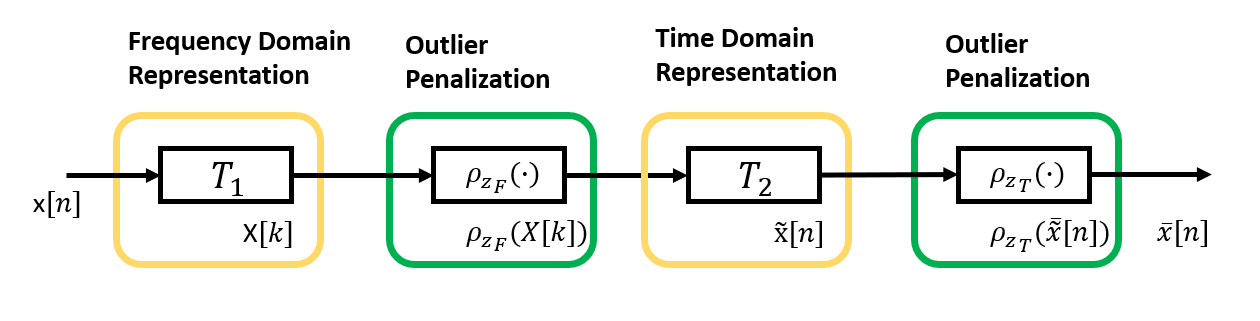}
	\caption{\ac{DD-RIM} with \ac{ZMNL}s applied at first in the frequency and then in the time domain.}
	\label{fig:Huber_DD_FT}
\end{figure}
As shown in Fig.~\ref{fig:Huber_DD_TF}, the received signal $x[n]$ is processed first with ${\rho_z}_{T}\left(\cdot\right)$ and then transformed into the frequency domain where ${\rho_z}_{F}\left(\cdot\right)$ is applied to get ${\rho_z}_F(X[k])$. Intuitively, the first nonlinearity would be in charge of mitigating outliers in the time domain (e.g. pulsed interferences) and the second nonlinearity of doing so in frequency (e.g. continuous wave interferences). The resulting cleaned signal, $\bar{x}[n]={\rho_z}_F(X[k])$ can be used to compute the robust CAF \eqref{DD_CAF} used for \ac{DD-RIM} \ac{DPE} processing as in \eqref{eq:MLEabs}.
Conversely, Fig.~\ref{fig:Huber_DD_FT} shows an alternative \ac{DD-RIM} configuration, where the nonlinearities order is changed. In Fig,~\ref{fig:Huber_DD_FT}, \ac{DD-RIM} approach is applied in the frequency domain first and then in the time domain. In detail, received signal $x[n]$ is transformed into the frequency domain signal, obtaining $X[k]$, which then is processed with \ac{RIM} technique ${\rho_z}_{F}\left(\cdot\right)$ and transformed back into the time domain where a second \ac{ZMNL} ${\rho_z}_{T}\left(\cdot\right)$ is applied.

\section{Loss of Efficiency}
\label{Sec:LoE}

In robust statistics, an importance performance metric is the so-called \textit{loss of efficiency}, or LoE for short. The \ac{LoE} is the performance of the estimator under nominal conditions, which in the context of this work is in the absence of an interference. The rationale is to quantify the degradation of the robust method, compared to the optimal method when the nominal conditions hold. Ideally, one would like that \ac{LoE} to be small.
%
Previous works on \ac{RIM} considered the output \ac{SNR} degradation in the absence of interferences for a number of \acp{ZMNL} \cite{borio2018huber}. Notice that in the \ac{DPE} case, this approach is less intuitive since the robust methodology is used to compute a position solution jointly processing satellite signals. Therefore, in this article we derive the \ac{LoE} for \ac{DPE}-\ac{RIM} in terms of its \ac{CRB} degradation, which we will show it is indeed related to the \ac{SNR} degradation of two-steps \ac{RIM}. To achieve that result, the section first presents the \ac{CRB} without \ac{RIM} and then, secondly, the achievable \ac{CRB} when \ac{RIM} is considered. In particular we focus on Huber's nonlinearity due to its superior performance \cite{borio2018huber,li2019dual}. transformations are considered. The \ac{LoE} is then established as the difference between the former \ac{CRB} and the \ac{RIM}-based solutions.

\subsection{Non-RIM}\label{sec:LOEnonRIM}
In order to obtain a more compact expression to compute the bound, we express \eqref{signalmodel} in vector form. Notice that, given that we are studying the \ac{LoE}, the interference $i[n]$ is not accounted for. The resulting signal model is
\begin{equation}
    \bm{x} = \bm{C}(\bkappa)\balpha + \bm{\eta},
\end{equation}
where $\balpha = [\alpha_1, \alpha_2, \cdots, \alpha_M]^\top \in \mathbb{C}^{M \times 1}$ is the complex-value amplitude vector of each signal; the $N$ signal samples are gathered in $\bm{x} = [x[0], \cdots, x[N-1]]^\top \in \mathbb{C}^{N \times 1}$ and $\bm{\eta}\in \mathbb{C}^{N \times 1}$ is a vector of $N$ AWGN samples, each drawn from $\mathcal{CN}(0, \sigma_n^2)$. $\bm{C} = (\bm{c}_1, \bm{c}_2, \cdots, \bm{c}_M) \in \mathbb{C}^{N \times M}$ is the joint local replica, in which each column is generated for corresponding satellites and each row is generated for different sampling instants.
More concretely, we have
\begin{equation}
\footnotesize
    \bm{c}_i = \left[\begin{matrix}
    s_i(-\tau_i(\bkappa))\\
    s_i(T_s-\tau_i(\bkappa))\mbox{e}^{(j2\pi f_{d,i}(\bkappa)T_s)}\\
    \vdots\\
    s_i((N-1)T_s-\tau_i(\bkappa))\mbox{e}^{(j2\pi f_{d,i}(\bkappa)(N-1)T_s)}
    \end{matrix}\right] \cred{= \left[\begin{matrix}
    \omega_{i,0}(\bkappa)\\
   \omega_{i,1}(\bkappa)\\
    \vdots\\
    \omega_{i,N-1}(\bkappa)
    \end{matrix}\right].}
\end{equation}
Considering this vector form for the received signal model, the log-likelihood function is proportional to 
\begin{equation}\label{eq:loglh3}
    \mathcal{L}(\bm{x}|\bkappa) =  - \frac{1}{\sigma_n^2}\left[\bm{x} - \bm{C}(\bkappa)\balpha\right]^H\left[\bm{x} - \bm{C}(\bkappa)\balpha\right]\;.
\end{equation}

Following the same derivation as in \cite{amigo2014cramer}, the \ac{FIM} is
\begin{equation}\label{eq:FIM2}
\begin{aligned}
    \bm{\mathcal{I}}(\bkappa) = 2\bm{P}^\top\bm{\Xi}\bm{\Gamma}\bm{P}. 
\end{aligned}
\end{equation}
$\bm{\Gamma} = \textrm{diag}(\bgamma^\top) = \textrm{diag}( [\textrm{SNR}_1, \textrm{SNR}_2, \cdots, \textrm{SNR}_M]^\top ) \in \mathbb{R}^{M \times M} = k_{\textrm{out}} \cdot \textrm{diag}( [\textrm{SNR}_1^{\textrm{out}}, \textrm{SNR}_2^{\textrm{out}}, \cdots, \textrm{SNR}_M^{\textrm{out}}]^\top ) \in \mathbb{R}^{M \times M}$ is the diagonal SNR matrix\cred{. $\bm{P}(\bkappa) = [\bm{P}^\top_1(\bkappa), \bm{P}^\top_2(\bkappa), \cdots, \bm{P}^\top_M(\bkappa)]^\top \in \mathbb{R}^{M \times 3}$ is the concatenation of $\bm{P}_i(\bkappa)$ from each satellite. The} $\textrm{SNR}_i$ denotes the prior-correlation SNR of the received signal from the $i$-th satellite
\begin{equation}\label{eq:SNR}
    \textrm{SNR}_i = \frac{1}{\sigma_n^2}\sum\limits_{n=0}^{N-1} \cred{\omega_{i,n}^*(\bkappa)\|\alpha_i\|^2\omega_{i,n}(\bkappa)} = \frac{1}{\sigma_n^2}\sum\limits_{n=0}^{N-1}\|\alpha_i\|^2s_i^2(\bkappa), 
\end{equation}
and $\textrm{SNR}_i^{\textrm{out}}$ denotes the corresponding post-correlation SNR while $ k_{\textrm{out}}$ is a scale parameter depending on the correlation form (i.e. correlation period, coherent correlation, noncoherent correlation).
The mean quadratic bandwidth (MQBD) of the signal, $\xi_i^2$, is defined as
\begin{equation}
    \xi_i^2 = \frac{\sum\limits_{n=0}^{N-1}s'^2_i[n]}{\sum\limits_{n=0}^{N-1}s^2_i[n]} = \frac{E_{s'}}{E_s} \;,
\end{equation}
\noindent such that $\bm{\Xi} = \textrm{diag}([\xi_1^2, \xi_2^2, \cdots, \xi_M^2]) \in \mathbb{R}^{M \times M}$ is the matrix form of MQBD. Typically, for a given GNSS constellation, a modulation scheme and a fixed bandwidth, the MQBD values are known and equal across signals of the same type. In other words, $\bm{\Xi} 
= \xi^2 \mathbf{I} 
\in \mathbb{R}^{M \times M}$ if the $M$ satellites are from the same constellation/signal. 
Recall that the inverse of the FIM in \eqref{eq:FIM2} provides the CRB for the parameters in $\bkappa$.

\subsection{RIM in Time Domain} \label{LoETD}
Comparing equation \eqref{eq:MLEabs} and \eqref{eq:robustMLEabs}, we can identify that the difference between those two solutions (i.e. non-RIM and RIM) is the \ac{ZMNL} $\rho_z(\cdot)$ applied to the received signal $x[n]$. Note that the same optimal solution for estimating $\btheta$ could be obtained when $x[n]$ is assumed as a heavy tailed distribution and when $\rho_z(x[n])$ is assumed to be Gaussian distributed. This section provides a CRB result under that assumption, which is then compared to the CRB in Section \ref{sec:LOEnonRIM} to quantify the LoE. 

In the previous section, we had a likelihood distribution of the form $\bm{x}|\bkappa \sim \mathcal{N}(\bm{C}(\bkappa)\balpha,\sigma_n^2 \mathbf{I})$. Once the nonlinearity is applied to the data, the resulting likelihood is derived in Appendix \ref{sec:ZMNLGuassian} as  $\rho_z(\bm{x})|\bkappa \sim \mathcal{N}(\bm{C}(\bkappa)\bar{\balpha},\bar{\sigma}_n^2 \mathbf{I})$,  
where $\bar{\balpha}$ is the distorted signal amplitude after the preprocessing and $\bar{\sigma}_n^2$ is the modified noise variance, related to the original parameters by \cite{borio2018huber, borio2018huberm}:
\begin{eqnarray}
		  \bar{\balpha} &=& \balpha\left[1 - \mbox{e}^{-\frac{T_h^2}{2\sigma_n^2}} + \frac{\sqrt{\pi}}{2}\frac{T_h}{\sqrt{2}\sigma_n}\mbox{erfc}\left(\frac{T_h}{\sqrt{2}\sigma_n}\right)\right]
\label{eq:time_amp_tilde} \\
	\bar{\sigma}_n^2 &=& \sigma_n^2\left[1 - \mbox{e}^{-\frac{T_h^2}{2\sigma_n^2}}\right].
\label{eq:time_var_tilde}
\end{eqnarray}
\noindent when the ZMNL is Huber's (cf. Appendix \ref{sec:RIMZMNL}), with this relation changing depending on the class of nonlinearity.

Under the assumed Gaussian model after applying the \ac{ZMNL} to the data, the corresponding FIM is:
\begin{equation}\label{eq:FIMrho}
    \bm{\mathcal{I}}_\rho(\bkappa) = 2\bm{P}^\top{\bm{\Xi}}\bm{\bar{\Gamma}}\bm{P}\;,
\end{equation}
\noindent where $\bar{\bm{\Gamma}} = \textrm{diag}(\bar{\bgamma}) = \textrm{diag}( [\bar{\textrm{SNR}}_1, \cdots, \bar{\textrm{SNR}}_M]^\top ) = k_{\textrm{out}} \cdot \textrm{diag}( [\bar{\textrm{SNR}}_1^{\textrm{out}}, \cdots, \bar{\textrm{SNR}}_M^{\textrm{out}}]^\top ) \in \mathbb{R}^{M \times M}$ is composed of the SNRs of the satellites computed as in \eqref{eq:SNR}, but with the modified parameters in \eqref{eq:time_amp_tilde} and \eqref{eq:time_var_tilde}.
Therefore, the \ac{CRB} after Huber's nonlinearity is applied to $x[n]$, in the time domain, would be given by $\bm{\mathcal{I}}_\rho^{-1}(\bkappa)$.

In order to define the LoE of the robust method, we consider the losses in \eqref{eq:time_amp_tilde} and \eqref{eq:time_var_tilde} impact on the post-correlation SNR of each satellite as a reduction by $L(\sigma_n, T_h) = \frac{\bar{\textrm{SNR}}_i^{\textrm{out}}}{\textrm{SNR}_i^{\textrm{out}}},  i\in \{1, \dots, M\}$.
As a consequence, it is easy to see that 
$\bm{\mathcal{I}}^{-1}(\bkappa) = \bm{\mathcal{I}_\rho}^{-1}(\bkappa) \cdot L(\sigma_n, T_h) $.

\subsection{RIM in Transformed Domain}\label{LoEFD}
Following the processing chain in Fig.~\ref{fig:genproscheme}, we have $\bar{x}[n]$ as the output signal when the \ac{ZMNL} function is applied in the transformed domain (in this case, the frequency domain which is the most common transformed domain in GNSS) \cite{borio2019robust}. Given the fact that $\mathbf{T}_1$ is a linear transformation, $X[k]$ is still Gaussian with expected value $\mathbb{E}\{X[k]\}$ and variance $\text{Var}\{X[k]\}$.
For instance, after the Huber's nonlinearity ${\rho_z}(\cdot)$, the mean and covariance of the resulting variable are modified as\cite{borio2018huber}:
\begin{equation}
		 \mathbb{E}\{{\rho_z}(X[k])\} = \mathbb{E}\{X[k]\}\left[1 - \mbox{e}^{-\frac{T_h^2}{2\sigma_n^2}} + \frac{\sqrt{\pi}}{2}\frac{T_h}{\sqrt{2}\sigma_n}\mbox{erfc}\left(\frac{T_h}{\sqrt{2}\sigma_n}\right)\right]
\label{eq:fistmomFinalfd}
\end{equation}
\cred{
\begin{equation}
	\text{Var}\{{\rho_z}(X[k])\} = \text{Var}\{X[k]\}\left[1 - \mbox{e}^{-\frac{T_h^2}{2\sigma_n^2}}\right].
\label{eq:variancefd}
\end{equation}}

When transforming the signal back to time domain, through the use of the linear transformation $\mathbf{T}_2$, several frequency samples ${\rho_z}(X[k])$ are combined to form the different time samples, $\bar{x}[n]$. By virtue of the \ac{CLT},  \cite{casella2021statistical} the resulting time domain signal $\bar{x}[n]$ can be considered to follow a Gaussian distribution as well \cite{borio2019robust}. Considering that $\mathbf{T}_2$ is also linear transform and $\mathbf{T}_1 \cdot \mathbf{T}_2 = \mathbf{I}$ is the identity operator (this holds for instance for FFT/IFFT operators), it was shown that $\bar{\bm{x}}|\bkappa \sim \mathcal{N}(\bm{C}(\bkappa)\bar{\balpha},\bar{\sigma}_n^2 \mathbf{I})$ has the same expected and variance values as in (\ref{eq:time_amp_tilde}) and (\ref{eq:time_var_tilde}). 
Following the same procedure as in earlier subsections, the \ac{FIM} and \ac{CRB} expressions can be obtained, respectively, as $\bm{I}_\rho(\bkappa) = 2\bm{P}^\top{\bm{\Xi}}\bar{\bm{\Gamma}}\bm{P}$ and $\textbf{CRB}_\rho(\bkappa) = \bm{I}_\rho^{-1}(\bkappa)$.


\subsection{RIM in Dual Domain}
Given the Gaussian assumption in \ac{RIM} time domain processing and the \ac{CLT} in \ac{RIM} transformed domain processing, we can assume our processed signal $\bar{x}[n]$ as Gaussian distribution after \ac{RIM} at \ac{DD} following a similar derivation as in earlier  subsections \cite{li2019dual}. Therefore, the log-likelihood of processed signal  after \ac{RIM} \ac{DD} processing is $\bar{\bm{x}}|\bkappa \sim \mathcal{N}(\bm{C}(\bkappa)\bar{\balpha},\bar{\sigma}_n^2 \mathbf{I})$
where 
\begin{equation}
		  \bar{\balpha} = \balpha\left[1 - \mbox{e}^{-\frac{T_h^2}{2\sigma_n^2}} + \frac{\sqrt{\pi}}{2}\frac{T_h}{\sqrt{2}\sigma_n}\mbox{erfc}\left(\frac{T_h}{\sqrt{2}\sigma_n}\right)\right]^2
\label{eq:fistmomFinaldd}
\end{equation}
\cred{
\begin{equation}
	\bar{\sigma}_n^2 = \sigma_n^2\left[1 - \mbox{e}^{-\frac{T_h^2}{2\sigma_n^2}}\right]^2.
\label{eq:variancefd2}
\end{equation}}
Following the same procedure, we can derive the \ac{FIM} and \ac{CRB} as $\bm{I}_\rho(\bkappa) = 2\bm{P}^\top{\bm{\Xi}}\bar{\bm{\Gamma}}\bm{P}$ and $\textbf{CRB}_\rho(\bkappa) = \bm{I}_\rho^{-1}(\bkappa)$, respectively.
$\bar{\text{SNR}}_i$ represents the updated SNR of $i$-th satellite signal under influence of RIM method in frequency domain, and $\bar{\bm{\Gamma}}$, the corresponding SNR matrix.

\section{Results}
\label{Sec:SimSetup}
\begin{figure}
\centering  \includegraphics[width=1.00\columnwidth]{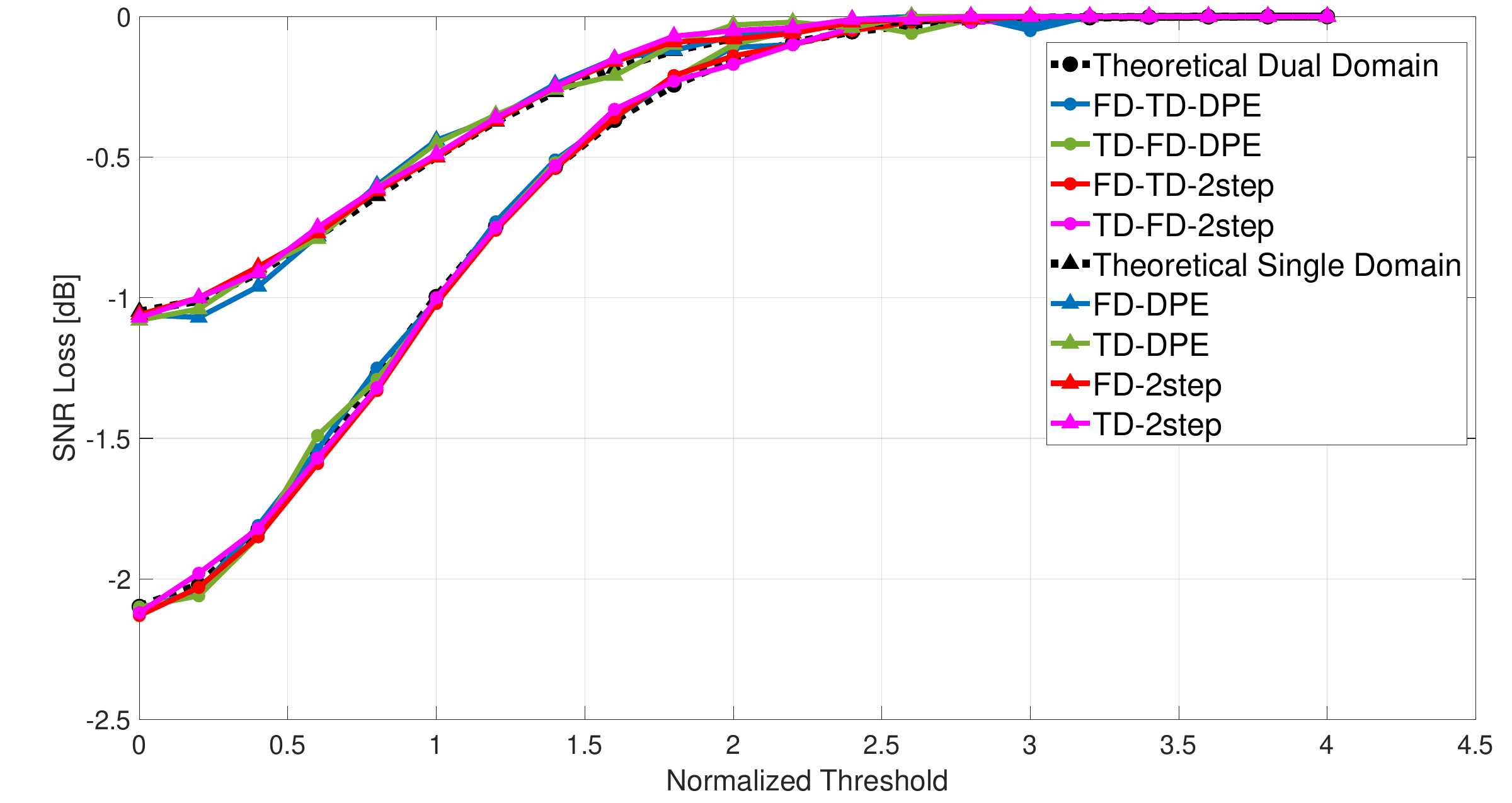}
  \caption{\ac{LoE} calculated from \ac{RMSE} of position estimation under different \ac{RIM} processing schemes.}
\label{fig::LoE_Simulation}
\end{figure}

Different experiments were run in order to validate the propose RIM DPE methodology. In particular, we first assessed the theoretical \ac{LoE} of the different \ac{RIM} flavours by a simulation of I\&Q samples from $7$ GPS L1 C/A satellites. In this experiment, the SNR of each transmitted signal was set to be the same, with a sampling frequency of $f_s = 50$ MHz and a frontend low-pass filter of $2$ MHz bandwidth. The receiver was simulated to be still at a fixed location. The LoE was computed by comparing the increase of \ac{RMSE} as a function of the \ac{CN0} when \ac{RIM} approaches are applied in the standard case (that is, when RIM processing is not applied). Without loss of generality, in order to avoid numerical errors, we conducted the LoE experiments under a moderately high $\ac{CN0}$ of $44$ dB-Hz for \ac{DPE} and $50$ dB-Hz for \ac{2SP} method. 
In these simulations, the \ac{ARS} numerical optimization method was employed to optimize \ac{DPE} cost function and estimate $\bkappa$ \cite{closas2009bayesian}. In the \ac{2SP} method, a \ac{LS} method was used to estimate $\bkappa$ using the pseudoranges produced by a \ac{CAF} maximization. The \ac{RMSE} is computed after averaging $5\cdot 10^4$ independent Monte Carlo experiments. 
Fig.~\ref{fig::LoE_Simulation} compares the LoE of various \ac{RIM} approaches, both for \ac{DPE} and \ac{2SP} methods, as a function of the normalized threshold $T_h$, an important parameter in Huber's non-linearity. In the figure, the black dashed lines represents the theoretical \ac{LoE} of both single domain \ac{RIM} (i.e. either time or frequency) and \ac{DD-RIM} approaches, where the line with circle represents \ac{DD-RIM} approaches and the lines with triangles indicate single domain \ac{RIM}. Similarly, the solid lines with circle also represent the experimentally computed \ac{LoE} of \ac{DD-RIM} approaches while those with triangles indicate experimental \ac{LoE} of single domain \ac{RIM} approaches. It can be observed that both DPE and 2SP approaches share the same LoE, given a \ac{RIM} processing scheme. Overall, the results should good agreement between theoretical and experimental LoE, thus validating our LoE derivation.

Another set of experiments were performed in order to assess the robustness of RIM-DPE. In particular, simulations considering both CW and DME interferences were tested, which are discussed here. \cred{The strength of the interference was adjusted with the \ac{JN}, defined as $\text{JN} = \frac{\alpha_I^2}{\sigma_n^2}$ with $\alpha_I$ being the amplitude of the interference.}
Similarly as before, a simulation of I\&Q samples from $7$ GPS \cred{L5} C/A satellites was generated, with $\ac{CN0} = 44 $ dB-Hz for all. The receiver employed a $20$ MHz bandwidth low-pass filter and was static throughout the experiment, which consisted of $50$ seconds worth of data. \cred{Note that the higher bandwidth is designed to include {GPS} L5. The threshold of Huber's ZMNL is chosen as $T_h = 1.345\hat{\sigma}_n$, which is generally picked to give reasonably high efficiency in the normal case, and the $\hat{\sigma}_n$ is calculated using the \ac{MAR} of received signal: $\hat{\sigma}_n =  \text{MAR}/0.6745$ \cite{fox2002robust}.}

\begin{figure}
\centering  \includegraphics[width=1.00\columnwidth]{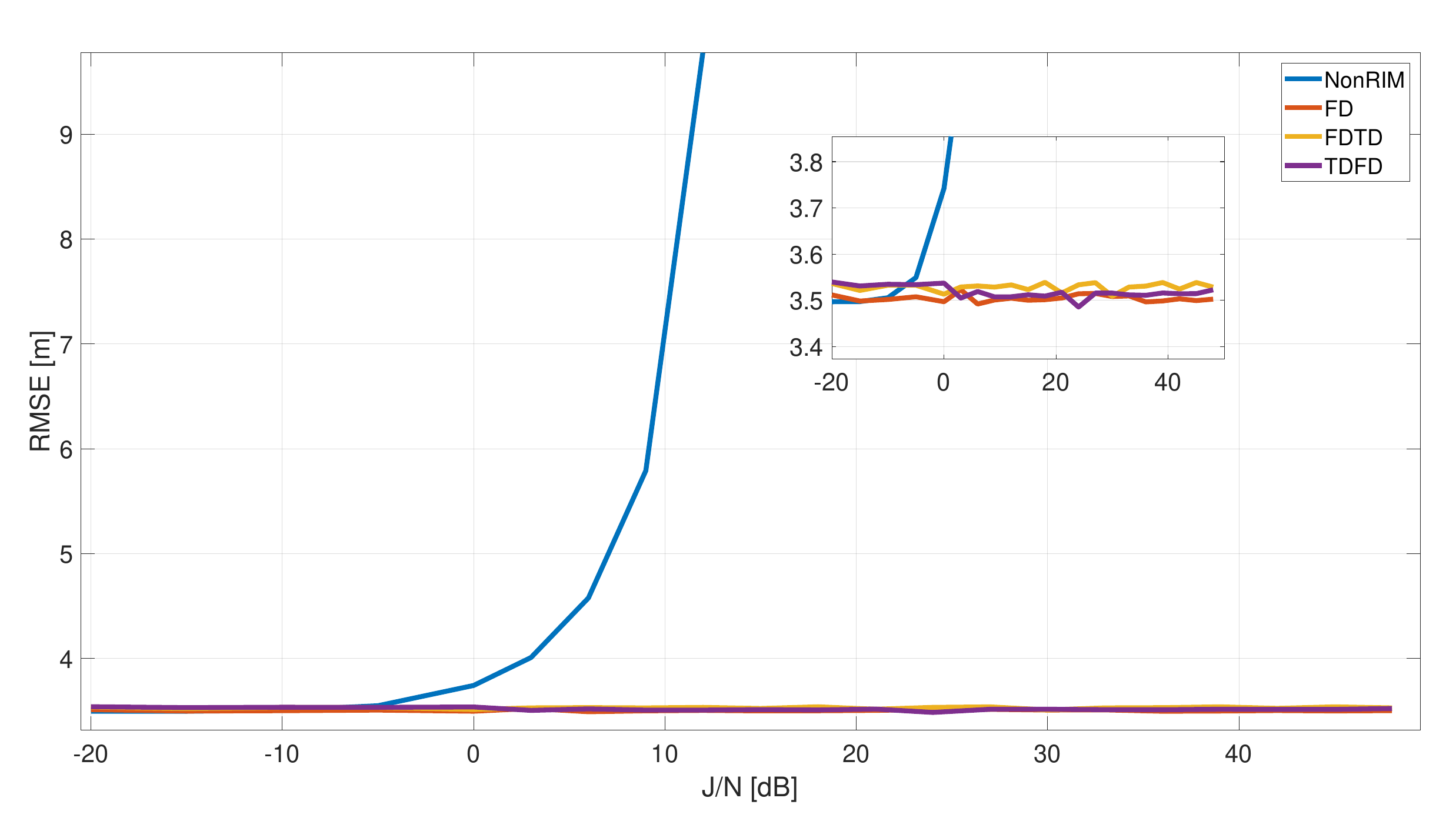}
  \caption{\ac{RMSE} of position estimation under different DPE \ac{RIM} processing techniques in the presence of a \ac{CW} jamming signal.}
\label{fig::RMSE_CW}
\end{figure}

\begin{figure}
\centering  \includegraphics[width=1.00\columnwidth]{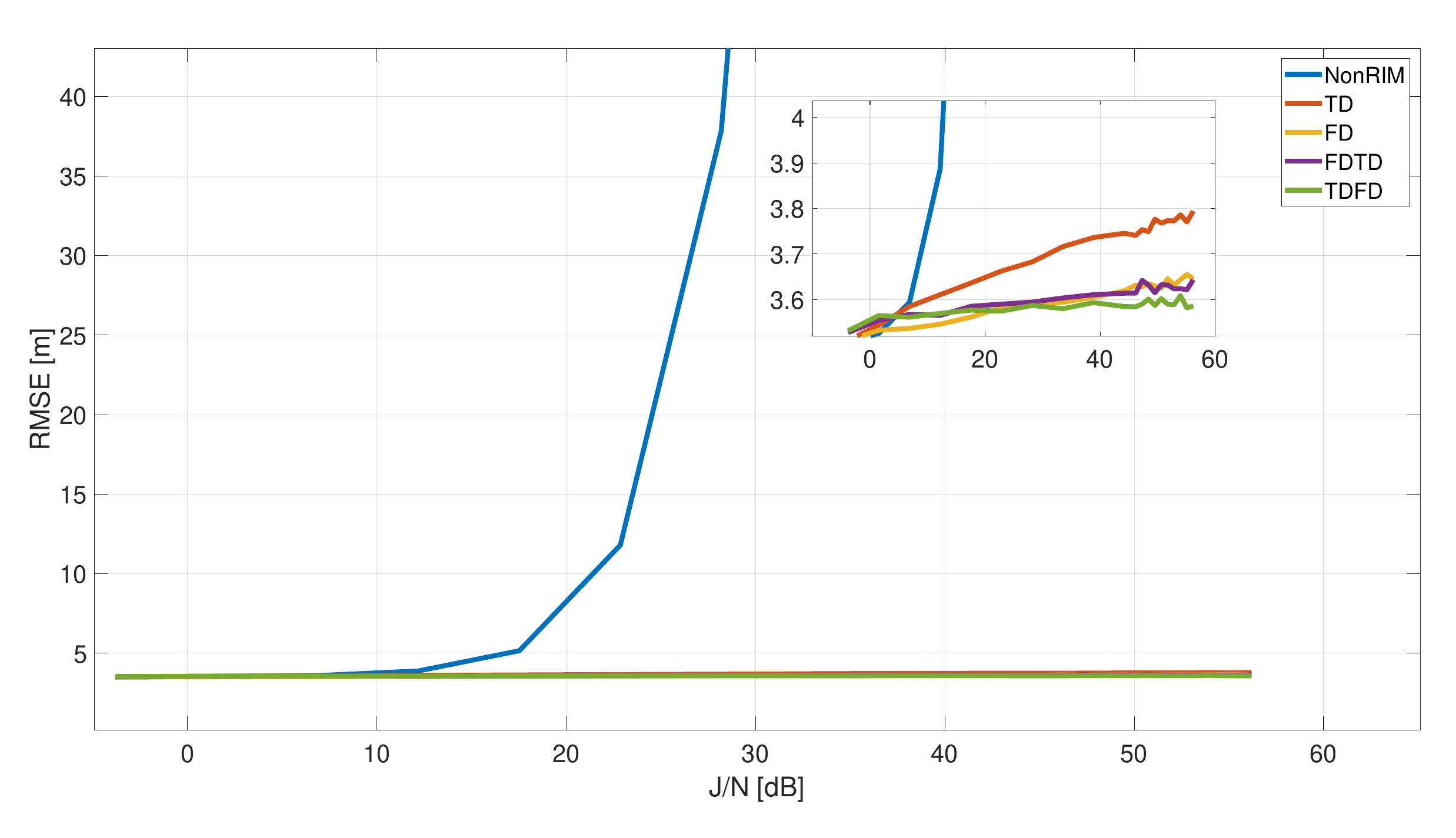}
  \caption{\ac{RMSE} of position estimation under different DPE \ac{RIM} processing techniques in the presence of a \ac{DME} interference signal.}
\label{fig::RMSE_DME}
\end{figure}

Fig. \ref{fig::RMSE_CW} shows the various \ac{RIM} approaches compared with the standard DPE non-RIM processing in the presence of a \ac{CW} jamming signal\cred{, with \ac{JN} varying from $-20$ dB to $48$ dB}. It can be observed that the case when \ac{RIM} is not used, the presence of a \ac{CW} jamming signal noticeably affects the estimation performance. In contrast, when DPE is used in conjunction with \ac{RIM} processing techniques, the results show relatively stable performances over different \ac{CW} power values. From the figure, we note that the best
performance is achieved when a single non-linearity is applied directly in the frequency domain. This result is consistent to previous works considering two-steps processing \cite{borio2018huber}, where it was noted that \ac{CW}s are maximally concentrated in the frequency domain. Nevertheless, results show that the use of DD-RIM does not significantly degrade interference mitigation performance. 
Similarly, Fig. \ref{fig::RMSE_DME} shows the performance of the same set of DPE approaches, in this case under the presence of a \ac{DME} interference signal\cred{, with \ac{JN} varying from $-4$ dB to $56$ dB}. The received DME power was modeled in the simulations considering the \ac{FSPLM}. In general, RIM in one domain was not able to effectively mitigate DME signals and DD-RIM provides the best performance as compared with single-domain RIM techniques. The most effective approach was obtained when time-then-frequency domain processing was implemented. More specifically, time-then-frequency domain processing performed better than
frequency-then-time domain processing. In the former, after time processing, the resulting signal is still relatively
sparse in the frequency domain and thus it can be further mitigated using a robust non-linearity. In the second case, frequency processing does not produce an interfering signal that is sparse in time and that can be exploited by RIM
in that domain. This ordering of RIM solutions is, again, consistent with the results for two-steps positioning reported in \cite{borio2018huber}.

\section{Conclusions}
\label{Sec:Conclusions}
Interference mitigation is crucial to protect GNSS from both intentional or unintentional interference signals. This paper presented the use of different \ac{RIM} approaches within a direct-positioning framework. \cred{RIM has the desirable feature of avoiding the estimation of the interference signal, thus simplifying its implementation when compared to interference cancellation methods.} \cred{Incorporating RIM} augments the range of applicability of \ac{DPE} in interference-rich situations, while DPE is already known to enhance the sensitivity of GNSS receivers to operate under weak signal conditions. The RIM methodology leverages results in robust statistics to design a new cross-ambiguity function and, consequently, a novel DPE cost function. In particular, this article explored the use of Huber’s non-linearity for complex-valued signals, showing remarkable performance results under \ac{CW} and \ac{DME} interferences. 
\cred{Notice that RIM, which DPE-RIM is based on, is effective against interference signals that can be considered to be outliers in time and/or transformed domains, which encompass most of the known GNSS interference threats, although not all.}
This paper
provided analytical expressions for the LoE of DPE RIM, that is, the degradation of performance
caused by the proposed robust methods under nominal conditions when the interference signal is not present,
showing negligible losses. 
\cred{DPE is a receiver framework that is known to provide enhanced sensitivity, enabling GNSS use in contested environments featuring weak signal conditions. The use of RIM in conjunction to DPE enables the high-sensitivity operation even under interference conditions. Future developments of DPE might involve its extension to high-accuracy applications.}


\appendices

\section{Parameter estimation under RIM processing}\label{sec:RIMDPE}
This appendix provides the derivation of the estimator for $\bkappa$ and amplitudes $\alpha_1,\dots,\alpha_M$ under RIM processing in a transformed domain, which results in the optimization of a robust version of the \ac{CAF}. To achieve this goal, we take a twofold process. First, we linearize the general cost function using a first-order Taylor, as was done earlier in \ac{2SP} works. This is explained in equations (\ref{Taylor series}) to (\ref{Transform formula}). Secondly, to estimate the additional amplitude parameters $\alpha_1,\dots,\alpha_M$, approximations based on the non-linearity are required, as derived in equations (\ref{h_approx1}) and (\ref{h_approx2}).
Then, both results are combined in order to obtain a general robust \ac{CAF} whose maximization would result in the \ac{RIM} solution.

In \ac{RIM} processing, the standard square error function is replaced by other choices that are able to attenuate the effect of model outliers. See Appendix \ref{sec:RIMZMNL} for an overview of those considered in the GNSS context of interest in this paper. Generally, the cost function to minimize under M-estimation framework is:
\begin{equation} 
\small
\begin{aligned}
&J_{\rho}(\bkappa) =\sum_{k=0}^{N-1}(\rho(\mathbf{T}_1(x[k] - \sum_{i=1}^{M}\alpha_ic_i(kT_{s}-\tau_{i}(\bkappa))\mbox{e}^{j(2\pi f_{d,i}(\bkappa)nT_{s}+\phi_i)}))) \\
&= \sum_{n=0}^{N-1} {\rho}((\mathbf{T}_1(x[k]) - \mathbf{T}_1(\sum_{i=1}^{M}\alpha_ic_i(kT_{s}-\tau_{i}(\bkappa))\mbox{e}^{j(2\pi f_{d,i}(\bkappa)kT_{s}+\phi_i)}))
\label{Cost function}
\end{aligned}
\end{equation}
where $\rho(\cdot)$ is a cost function, which is a design choice that depends on the modeling assumptions. For instance, if $\rho(\cdot)$ is $\left| \cdot \right|^2$, we obtain the standard least squares solution, as shown in \eqref{eq:MLEabs}. $\mathbf{T}_1$ is the linear transforms defined in Section ~\ref{Sec:RIM}-\ref{sec:TDRIM}. { {Note that $\mathbf{T}_1$ is a unitary matrix, satisfying $\mathbf{T}_1 \circ \mathbf{T}_1^H = \mathbf{I}$. In other words, $\mathbf{T}_2  = \mathbf{T}_1^H$.  }}
According to the fact that received GNSS signals are weak and the
signal amplitude $\alpha_i$ can be assumed to be small compared to the noise term, $\rho(\cdot)$ can be expanded in Taylor series \cite{borio2017myriad} for small amplitudes. Function $\rho(\cdot)$ can be regarded as a real function of two real variables, the real and imaginary parts of the complex signal. That is, with $z\in\mathbb{C}$ we can express:
\begin{equation} 
\rho(z)=\rho(z_{I},z_{Q})
\label{rho_general}
\end{equation}
which, for a small increment $\Delta z=\Delta z_{I}+j\Delta z_{Q}$, can be expressed as
\begin{equation} 
\begin{aligned}
\rho(z-\Delta z)&=\rho(z_{I}-\Delta z_{I},z_{Q}-\Delta z_{Q})\\
&\backsimeq \rho(z)-\frac{\partial \rho(z)}{\partial z_{I}} \Delta z_{I}-\frac{\partial \rho(z)}{\partial z_{Q}} \Delta z_{Q}\\
&=\rho(z)-{\Re}\{\rho_z(z)\Delta z^*\}\label{Taylor series}
\end{aligned}
\end{equation}
where $z^\ast$ denotes complex conjugate of $z$ and 
\begin{equation} 
\rho_z(z)=\rho_I(z)+j\rho_Q(z)=\frac{\partial \rho(z)}{\partial z_{I}}+j\frac{\partial \rho(z)}{\partial z_{Q}} = 2\frac{\partial \rho(z)}{\partial z^*}
\label{Transform formula}
\end{equation}
\begin{equation} 
\rho_{\bar{z}}(z)=\rho_I(z)-j\rho_Q(z)=\frac{\partial \rho(z)}{\partial z_{I}}-j\frac{\partial \rho(z)}{\partial z_{Q}} = 2\frac{\partial \rho(z)}{\partial z}
\label{Transform formula}
\end{equation}
According to \eqref{Taylor series}, \eqref{Cost function} can be approximated as
\begin{equation} 
\small
\begin{aligned}
J_{\rho}(\bkappa)&\backsimeq \sum_{n=0}^{N-1} \rho(\mathbf{T}_1(x[k])) \\
& - {\Re} \{\sum_{k=0}^{N-1}{\rho}_z(\mathbf{T}_1(x[k]))\\
&\mathbf{T}_1(\sum_{i=1}^{M}\alpha_ic_i(kT_{s}-\tau_{i}(\bkappa))\mbox{e}^{j(2\pi f_{d,i}(\bkappa)kT_{s}+\phi_i)})^*\}\\
\label{Transformed cost function}
\end{aligned}
\end{equation}
Since the first term in \eqref{Transformed cost function} does not depend on the parameter 
$\bkappa$, minimizing the cost function could be transformed into maximizing
\begin{equation}
\small
\begin{aligned}
 J_{\textrm{real}}(\bkappa) 
   & = {\Re} \{\sum_{k=0}^{N-1}{\rho}_z(\mathbf{T}_1(x[k]))\mathbf{T}_1(\sum_{i=1}^{M}\alpha_ic_i(kT_{s}-\tau_{i}(\bkappa))\mbox{e}^{j(2\pi f_{d,i}(\bkappa)kT_{s}+\phi_i)})^*\}\\
   & \propto  {\Re} \{\sum_{n=0}^{N-1}\mathbf{T}_2({\rho}_z(\mathbf{T}_1(x[n]))) \\
   &\mathbf{T}_2(\mathbf{T}_1(\sum_{i=1}^{M}\alpha_ic_i(nT_{s}-\tau_{i}(\bkappa))\mbox{e}^{j(2\pi f_{d,i}(\bkappa)nT_{s}+\phi_i)}))^*\}\\
 & = {\Re} \{\sum_{n=0}^{N-1} \Tilde{\rho}_z(x[n])\sum_{i=1}^{M}\alpha_ic_i(nT_{s}-\tau_{i}(\bkappa))\mbox{e}^{-j(2\pi f_{d,i}(\bkappa)nT_{s}+\phi_i)}\}\\
 & = {\Re} \{\sum_{i=1}^{M}\sum_{n=0}^{N-1}\Tilde{\rho}_z(x[n])\alpha_ic_i(nT_{s}-\tau_{i}(\bkappa))\mbox{e}^{-j(2\pi f_{d,i}(\bkappa)nT_{s}+\phi_i)}\}\\
  & = \sum_{i=1}^{M}\alpha_i{\Re} \{\sum_{n=0}^{N-1}\Tilde{\rho}_z(x[n])c_i(nT_{s}-\tau_{i}(\bkappa))\mbox{e}^{-j(2\pi f_{d,i}(\bkappa)nT_{s}+\phi_i)}\}
 \label{realcost}
\end{aligned}
\end{equation}
which is a function of both $\bkappa$ and the amplitudes $\alpha_1,\dots,\alpha_M$, and $\Tilde{\rho}_z(x[n]) = \mathbf{T}_2({\rho}_z(\mathbf{T}_1(x[n])))$.The proportional symbol in the equation above comes from the Parseval's theorem and the fact that $\mathbf{T}_1$ and $\mathbf{T}_2$ are linear and can be represented as unitary matrix.
To achieve the optimal estimation of $\bkappa$, we first need to estimate the $\alpha_i$:
\begin{equation}
\footnotesize
\begin{aligned}
     \hat{\alpha}_i &= \argmin_{\alpha_i}\sum_{k=0}^{N-1}{\rho}(\mathbf{T}_1(x[k] - \sum_{i=1}^{M}\alpha_ic_i(kT_{s}-\tau_{i}(\bkappa))\mbox{e}^{j(2\pi f_{d,i}(\bkappa)kT_{s}+\phi_i)}))\\
     &= \argmin_{\alpha_i} \sum_{k=0}^{N-1} {\rho}((\mathbf{T}_1(x[k]) - \mathbf{T}_1(\sum_{i=1}^{M}\alpha_ic_i(kT_{s}-\tau_{i}(\bkappa))\mbox{e}^{j(2\pi f_{d,i}(\bkappa)kT_{s}+\phi_i)}))
\end{aligned}
\end{equation}
whose derivative with respect to $\alpha_i$ is (following the chain rule):
\begin{equation}
\footnotesize
\begin{aligned}
     &\Re \{2\sum_{k=0}^{N-1} \mathbf{T}_1(c_i(kT_{s}-\tau_{i}(\bkappa))\mbox{e}^{j(2\pi f_{d,i}(\bkappa)kT_{s}+\phi_i)}) \\
     &\rho_{\bar{z}}(\mathbf{T}_1(x[k] - \alpha_ic_i(kT_{s}-\tau_{i}(\bkappa))\mbox{e}^{j(2\pi f_{d,i}(\bkappa)kT_{s}+\phi_i)}))\} = 0
     \label{alphaiest}
\end{aligned} 
\end{equation}
The equation above can be further simplified when one accounts for the properties of the most common \acp{ZMNL} used in \ac{RIM} processing, as reviewed in Appendix \ref{sec:RIMZMNL}.
For instance, we can identify that the cost functions are all functions of the absolute value of a sample. Therefore we can further express \eqref{rho_general} as:
\begin{equation}\label{eq:abszreq}
    \rho(z) = g(|z|)
\end{equation}
with first derivative 
\begin{equation}
\footnotesize
    \rho_z(z) = 2\frac{\partial g(|z|)}{\partial |z|} \frac{\partial |z|}{\partial z^*} = 2 \frac{\partial g(|z|)}{\partial |z|} z \triangleq  \frac{z}{h(|z|)}  \label{rhozderive}
\end{equation}
where $h(|z|) \triangleq  \frac{1}{2 \frac{\partial g(|z|)}{\partial |z|}}$,
and similarly we have:
\begin{equation}
\footnotesize
    \rho_{\bar{z}}(z) = 2\frac{\partial g(|z|)}{\partial |z|} \frac{\partial |z|}{\partial z} = 2 \frac{\partial g(|z|)}{\partial |z|} z^* \triangleq  \frac{z^*}{h(|z|)}  \label{rhozderiveconjugate}
\end{equation}
It can be seen that \eqref{eq:abszreq}is satisfied by the common \ac{ZMNL} choices, cf. Appendix~\ref{sec:RIMZMNL}, for instance observing that the different $\rho(\cdot)$ are a function of the magnitude of the argument.
Using \eqref{rhozderiveconjugate}, we have:
\begin{equation}
\footnotesize
\begin{aligned}
     &\Re \{2\sum_{k=0}^{N-1} \mathbf{T}_1(c_i(kT_{s}-\tau_{i}(\bkappa))\mbox{e}^{j(2\pi f_{d,i}(\bkappa)kT_{s}+\phi_i)})\\ & \frac{ \mathbf{T}_1(x[k] - \alpha_ic_i(kT_{s}-\tau_{i}(\bkappa))\mbox{e}^{j(2\pi f_{d,i}(\bkappa)kT_{s}+\phi_i)})^* }{h(|\mathbf{T}_1(x[k] - \alpha_ic_i(kT_{s}-\tau_{i}(\bkappa))\mbox{e}^{j(2\pi f_{d,i}(\bkappa)kT_{s}+\phi_i)}) |)} \}= 0
\end{aligned} 
\end{equation}
Given that $\alpha_i$ is relatively small compared to $x[n]$, we have an approximation of the denominator term in  as:
\begin{equation}
\footnotesize
\begin{aligned}
      &\Re \{2\sum_{n=0}^{N-1} \mathbf{T}_1(c_i(nT_{s}-\tau_{i}(\bkappa))\mbox{e}^{j(2\pi f_{d,i}(\bkappa)nT_{s}+\phi_i)})\\ & \frac{ \mathbf{T}_1(x[n] - \alpha_ic_i(nT_{s}-\tau_{i}(\bkappa))\mbox{e}^{j(2\pi f_{d,i}(\bkappa)nT_{s}+\phi_i)})^* }{h(|\mathbf{T}_1(x[n]) |)} \}= 0
      \label{h_approx1}
\end{aligned} 
\end{equation}
Considering the unitary property and linearity of the matrices corresponding to $\mathbf{T}_1$ and $\mathbf{T}_2$, as well as the Parseval's theorem, the equation above can be transformed as:
\begin{equation}
\footnotesize
\begin{aligned}
      &\Re \{2\sum_{n=0}^{N-1} \mathbf{T}_2(\mathbf{T}_1(c_i(nT_{s}-\tau_{i}(\bkappa))\mbox{e}^{j(2\pi f_{d,i}(\bkappa)nT_{s}+\phi_i)}))\\ & \mathbf{T}_2(\frac{ \mathbf{T}_1(x[n] - \alpha_ic_i(nT_{s}-\tau_{i}(\bkappa))\mbox{e}^{j(2\pi f_{d,i}(\bkappa)nT_{s}+\phi_i)}) }{h(|\mathbf{T}_1(x[n]) |)})^* \}= 0
\end{aligned} 
\end{equation}
leading to
\begin{equation}
\footnotesize
\begin{aligned}
     &\Re \{2 \sum_{n=0}^{N-1} c_i(nT_{s}-\tau_{i}(\bkappa))\mbox{e}^{j(2\pi f_{d,i}(\bkappa)nT_{s}+\phi_i)}\mathbf{T}_2(\frac{\mathbf{T}_1(x[n]) }{h(|\mathbf{T}_1(x[n]) |)} )^*\}
     \\
     &-\alpha_i \Re \{2 \sum_{n=0}^{N-1} c_i(nT_{s}-\tau_{i}(\bkappa))\mbox{e}^{j(2\pi f_{d,i}(\bkappa)nT_{s}+\phi_i)}\\
     &\mathbf{T}_2(\frac{\mathbf{T}_1(c_i(nT_{s}-\tau_{i}(\bkappa))\mbox{e}^{j(2\pi f_{d,i}(\bkappa)nT_{s}+\phi_i)})}{h(|\mathbf{T}_1(x[n]) |)} )^* \}
     \\
      &=\Re \{2 \sum_{n=0}^{N-1} c_i(nT_{s}-\tau_{i}(\bkappa))\mbox{e}^{j(2\pi f_{d,i}(\bkappa)nT_{s}+\phi_i)}\Tilde{\rho}_z(x[n])^*\}
     \\
     &-\alpha_i \Re \{2 \sum_{n=0}^{N-1} c_i(nT_{s}-\tau_{i}(\bkappa))\mbox{e}^{j(2\pi f_{d,i}(\bkappa)nT_{s}+\phi_i)}\\
     &\mathbf{T}_2(\frac{\mathbf{T}_1(c_i(nT_{s}-\tau_{i}(\bkappa))\mbox{e}^{j(2\pi f_{d,i}(\bkappa)nT_{s}+\phi_i)})}{h(|\mathbf{T}_1(x[n]) |)} )^* \} = 0
\end{aligned} 
\end{equation}
{
Given that \ac{RIM} processing is based on the assumption that the interference component occurs as sparse representation in the processed domain and few samples are affected, we have the assumption that $h(|\mathbf{T}_1(\bm{x})|) \propto  \mathbf{I}$, where $\bm{x}$ is a vector with $x[n]$ as its $n$-th element.
With this assumption, we have
\begin{equation}
    \begin{aligned}
   &\Re \{2 \sum_{n=0}^{N-1} c_i(nT_{s}-\tau_{i}(\bkappa))\mbox{e}^{j(2\pi f_{d,i}(\bkappa)nT_{s}+\phi_i)}\\
     &\mathbf{T}_2(\frac{\mathbf{T}_1(c_i(nT_{s}-\tau_{i}(\bkappa))\mbox{e}^{j(2\pi f_{d,i}(\bkappa)nT_{s}+\phi_i)})}{h(|\mathbf{T}_1(x[n]) |)} )^* \}  \propto 2N
    \end{aligned}
    \label{h_approx2}
\end{equation}
leading to 
\begin{equation}
\footnotesize
\begin{aligned}
\hat{\alpha}_i &\propto \Re \{ \sum_{n=0}^{N-1} c_i(nT_{s}-\tau_{i}(\bkappa))\mbox{e}^{j(2\pi f_{d,i}(\bkappa)nT_{s}+\phi_i)}\Tilde{\rho}_z(x[n])^*\}
\\
&= \Re \{ \sum_{n=0}^{N-1} c_i(nT_{s}-\tau_{i}(\bkappa))\mbox{e}^{-j(2\pi f_{d,i}(\bkappa)nT_{s}+\phi_i)}\Tilde{\rho}_z(x[n])\}
\label{alphaiestapprox}
\end{aligned} 
\end{equation}
}
Substituting \eqref{alphaiestapprox} into \eqref{realcost}, we have $J_{\textrm{real}}(\bkappa)$ as 
\begin{equation}
\footnotesize
\begin{aligned}
J_{\textrm{real}}(\bkappa) & \propto \sum_{i=1}^{M}{\Re} \{\sum_{n=0}^{N-1}\Tilde{\rho}_z(x[n])c_i(nT_{s}-\tau_{i}(\bkappa))\mbox{e}^{-j(2\pi f_{d,i}(\bkappa)nT_{s}+\phi_i)}\}^2\\
&=\sum_{i=1}^{M}{\Re} \{\mathcal{C}_{\rho,i}(\bkappa)\mbox{e}^{-j\phi_i}\}^2\\
&=\sum_{i=1}^{M}{\Re} \{\big|\mathcal{C}_{\rho,i}(\bkappa)\big|\mbox{e}^{j(\angle{\mathcal{C}_{\rho,i}(\bkappa)}-\phi_i)}\}^2\\
&=\sum_{i=1}^{M}\big(\big|\mathcal{C}_{\rho,i}(\bkappa)\big|{\Re} \{\mbox{e}^{j(\angle{\mathcal{C}_{\rho,i}(\bkappa)}-\phi_i)}\}\big)^2\\
&=\sum_{i=1}^{M}\big(\big|\mathcal{C}_{\rho,i}(\bkappa)\big| \cos(\angle{\mathcal{C}_{\rho,i}(\bkappa)}-\phi_i)\big)^2\\
\end{aligned}
\label{correlation1}
\end{equation}
According to \eqref{correlation1}, the cost function
is factored in two terms. The first is the
absolute value of the \ac{CAF} and depends only on $\kappa$. The
second term is a cosine which also depends on $\phi_i$. The cosine can
be maximised by setting
with $\hat{\phi}_i = \angle{\mathcal{C}_{\rho,i}(\bkappa)}$, we can further convert the optimization of \eqref{correlation1} to:
\begin{equation}
\begin{aligned}
\hat{\bkappa}
& = \argmax_{\bkappa}\sum_{i=1}^{M}|\mathcal{C}_{\rho, i}(\bkappa)|^2
\end{aligned}
\label{correlation2}
\end{equation}
where $\mathcal{C}_{\rho, i}(\bkappa)$ is the robust version of {CAF} define as
\begin{equation}
\mathcal{C}_{\rho, i}(\bkappa)=\Tilde{\rho}_z(x[n])c_i(nT_{s}-\tau_{i}(\bkappa))\mbox{e}^{-j2\pi f_{d,i}(\bkappa)nT_{s}} \;. \label{Robust CAF}
\end{equation}


\section{Selected Non-linearities for RIM processing}\label{sec:RIMZMNL}
This appendix provides an overview of some RIM non-linearities considered in the \ac{GNSS} context of interest.
In \ac{GNSS} signal processing, the most common cost functions $\rho(z)$  are introduced in \cite{huber1964robust,huber2011robust,boriognss}, among which three of the \ac{ZMNL}s as well as the corresponding cost functions are listed in this section as examples:

\textit{1) Laplacian model assumption for the likelihood distribution \cite{daniele2018complex}.} The cost function $\rho(z)$ is:
\begin{equation}
\rho(z)=|z| 
\label{laplace_cost}
\end{equation} 
Then, the \ac{ZMNL} function ${\rho_z}(z)$ in (\ref{eq:finalsample}) can be obtained as (\ref{Transform formula}):
\begin{equation}
\rho_z(z)=\frac{z}{|z|}\triangleq {\rho_z}(z) \qquad \mbox{for } z \neq 0
\label{laplace_zmnl}
\end{equation}
The \ac{ZMNL} in (\ref{laplace_zmnl}) is referred to as \emph{complex signum}  \ac{ZMNL} according to \cite{CSignum}.
Furthermore, we have that
\begin{equation}
    \rho_{zz}(z) = \frac{-z^2}{|z|^3} \qquad \mbox{for } z \neq 0 \;,
\end{equation}
as needed with the \ac{DPE} \ac{RIM} framework discussed in this paper.

\textit{2) Cauchy model assumption for the likelihood distribution \cite{borio2017myriad}.}
The cost function $\rho(z)$ is:
\begin{equation}
\rho(z)=\frac{3}{2}\log({ K_C}+|z|^2)+\frac{1}{2}\log(\frac{4\pi^2}{K_C})
\end{equation} 
where $K_C$ is referred to as the linearity parameter\cite{arce2005nonlinear}.
The corresponding \emph{myriad} \ac{ZMNL} is:
\begin{equation}
{\rho_z}(z)=\frac{{K_C}z}{{K_C}+|z|^2} \label{Cauchy_update}
\end{equation}
with
\begin{equation}
    \rho_{zz}(z) = \frac{-K_C z^2}{(K_C+|z|^2)^2}
\end{equation}

\textit{3) M-estimation based on Huber's loss \cite{borio2018huber}} 
The cost function $\rho(z)$ is defined as:
\begin{equation}
\rho(z) = \left\{
		\begin{array}{ll}
			\frac{1}{2}|z|^2 & \qquad \mbox{for } |z| \leq T_h\\
			 T_h|z|-\frac{1}{2}T_h^2& \qquad \mbox{for } |z| > T_h\\
		\end{array}
	\right.
	\label{Huber_cost}
\end{equation}
By using  (\ref{Transform formula}), the resulting \ac{ZMNL} ${\rho_z}(z)$ is:
\begin{equation}
	{\rho_z}(z) \triangleq \rho_z(z) = \left\{
		\begin{array}{ll}
			z & \qquad \mbox{for } |z| \leq T_h\\
			T_h \; \mbox{csign}(z) & \qquad \mbox{for } |z| > T_h\\
		\end{array}
	\right.
	\label{huber_cost_z}
\end{equation}
where $T_h$ is a decision threshold, that is a tuning constant\cite{huber2011robust}, and $\mbox{csign}(z)$ is defined as: 
\begin{equation}
\mbox{csign}(z) = \left\{
		\begin{array}{ll}
		\frac{z}{|z|} & \qquad \mbox{for } z \neq 0\\
		0 & \qquad \mbox{for } z = 0
		\end{array}
	\right. .
	\label{huber_csign}
\end{equation}
such that
\begin{equation}
\rho_{zz}(z)  = \left\{
		\begin{array}{ll}
			0 & \qquad \mbox{otherwise}  \\
			T_h \frac{-z^2}{|z|^3} & \qquad \mbox{for} |z| > T_h \mbox{ and } z \neq 0 \\
		\end{array}
	\right.
	\label{rhozzhuber}
\end{equation}

\section{Maximum likelihood estimation after RIM non-linearity}
\label{sec:ZMNLGuassian}
This appendix shows the derivation of maximum likelihood estimator of $\bkappa$ once the RIM nonlinearity is applied. The Gaussian model assumption is shown in subsection~\ref{Sec:LoE}.~\ref{LoETD}. To estimate $\bkappa$, the maximum likelihood estimator is applied in (\ref{cost1}):
\begin{equation}
\begin{aligned}
    \hat{\bkappa} &= \underset{\bkappa}{\arg\min} \; J_{\rho}(\bkappa)\\
    &= \underset{\bkappa}{\arg\min} \left[\rho_z(\bm{x}) - \bm{C}(\bkappa)\bar{\balpha}\right]^H\left[\rho_z(\bm{x}) - \bm{C}(\bkappa)\bar{\balpha}\right]
    \label{cost1}
\end{aligned}
\end{equation}
where $\bar{\balpha}$ is the distorted signal amplitude after the non-linearity processing, related to the original parameters by \cite{borio2018huber, borio2018huberm}.
To minimize the cost function, we first take derivative w.r.t. $\bar\balpha$ and setting it to zero yields to 
\begin{equation}
    \hat{\bar{\balpha}} = (\bm{C}^H \bm{C})^{-1}\bm{C}^H\rho_z(\bm{x}).
    \label{alphaest}
\end{equation}
which turns in to $\hat{\bar{\balpha}} = \bm{C}^H\rho_z(\bm{x})$, given the property that $\bm{C}^H\bm{C} \approx \bm{I}$ \cite{gao2007dme}.
Substituting equation \eqref{alphaest} into equation \eqref{cost1}, it can be seen that 
\begin{equation}
\begin{aligned}
    J_{\rho}(\bkappa) &=  \left[\rho_z(\bm{x}) - \bm{C}(\bzeta)\hat{\bar{\balpha}}\right]^H\left[\rho_z(\bm{x})- \bm{C}(\bzeta)\hat{\bar{\balpha}}\right] \\
    &= \| \bm{C}^H\rho_z(\bm{x}) \| ^2 =
 \mathcal{C}_{\rho, i}(\bkappa) \;.
\end{aligned}
\label{CAF_Gaussian_rhoz}
\end{equation}
which is the vector form of the robust \ac{CAF} in (\ref{eq:robcaf}). This equality shows that the Gaussian assumption on $\rho_z(x[n])$ leads to the same $\bkappa$ estimation as under the actual distribution, as shown in Appendix \ref{sec:RIMDPE}.
As a consequence, this modeling assumption can be used to derive the estimation bounds, which greatly simplifies the calculations. 
\bibliographystyle{IEEEtran}
\bibliography{IEEEbrv, main}

\begin{IEEEbiography}[{\includegraphics[width=1in,height=1.25in,clip,keepaspectratio]{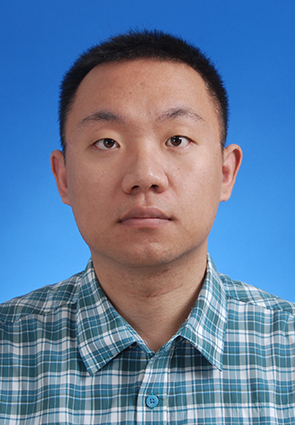}}]{Haoqing Li}{\space}received the B.S. degree in electrical engineering from Wuhan University, China, in 2016 and the M.S. degree in electrical and computer engineering from Northeastern University, Boston, MA, in 2018, where he is currently working toward the Ph.D. degree in electrical and computer engineering.
His research interests include GNSS signal processing, anti-jamming technology, and robust statistics.
\end{IEEEbiography}

\begin{IEEEbiography}[{\includegraphics[width=1in,height=1.25in,clip,keepaspectratio]{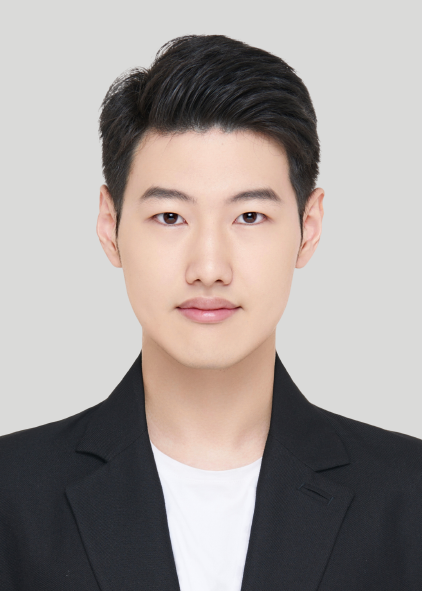}}]{Shuo Tang}{\space} received the B.S. degree in mechanical engineering from China Agricultural University, China and the M.S. degree in mechanical engineering from Northeastern University, Boston, MA, in 2014 and 2018, respectively.
He is currently working as a Ph.D. candidate in electrical and computer engineering at Northeastern University. His research interests include GNSS signal processing, sensor fusion and computational statistics.
\end{IEEEbiography}%

\begin{IEEEbiography}[{\includegraphics[width=1in,height=1.25in,clip,keepaspectratio]{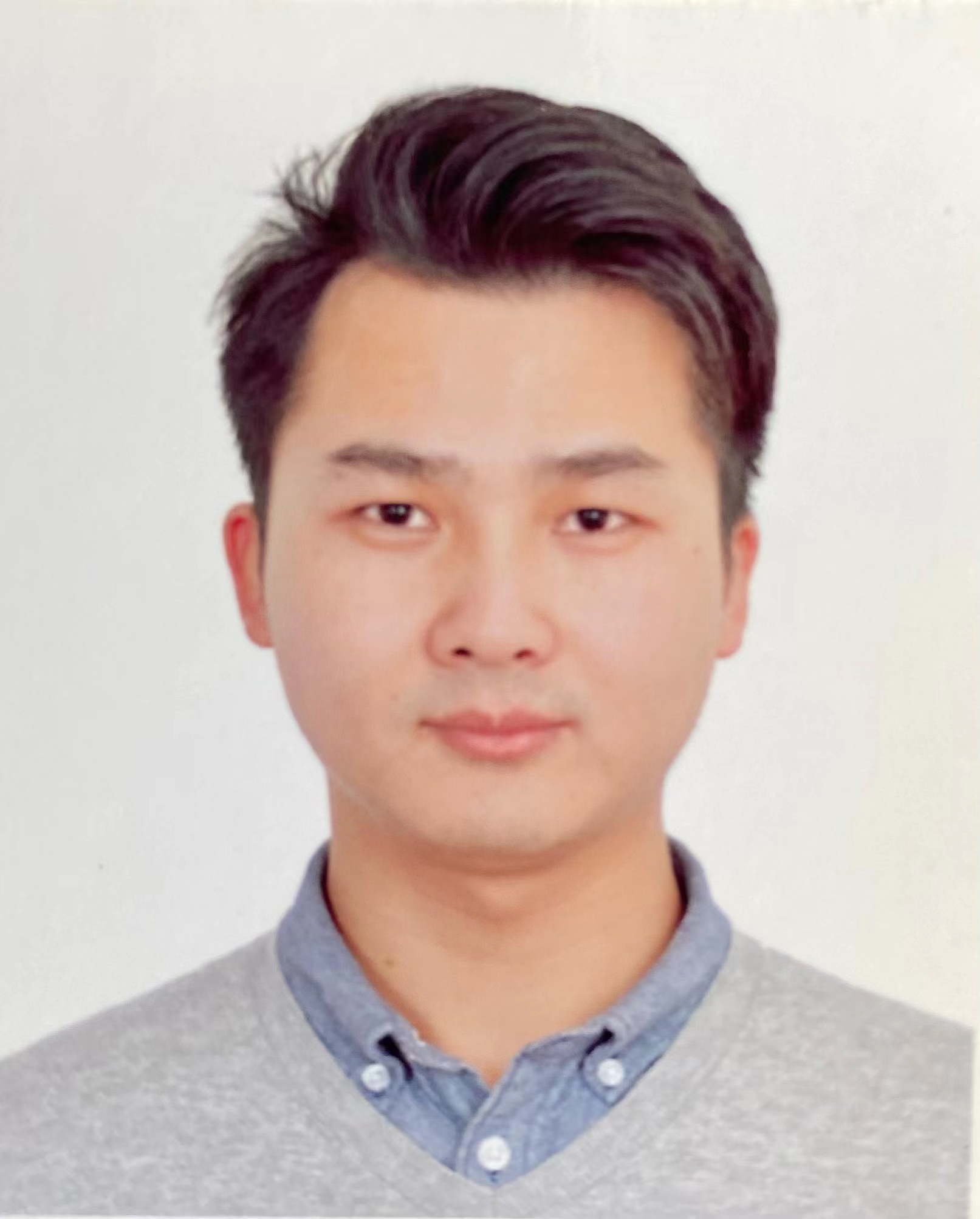}}]{Peng Wu}{\space}received his B.S. degree in Physics from Tianjin University of Technology, China and M.S. degree in Electrical Engineering from Northeastern University, Boston, MA. He is currently a Ph.D. candidate in the Department of Electrical and Computer Engineering at Northeastern University. His research interests include distributed data fusion and machine learning with applications to indoor positioning and tracking.
\end{IEEEbiography}

\begin{IEEEbiography}[{\includegraphics[width=1in,height=1.25in,clip,keepaspectratio]{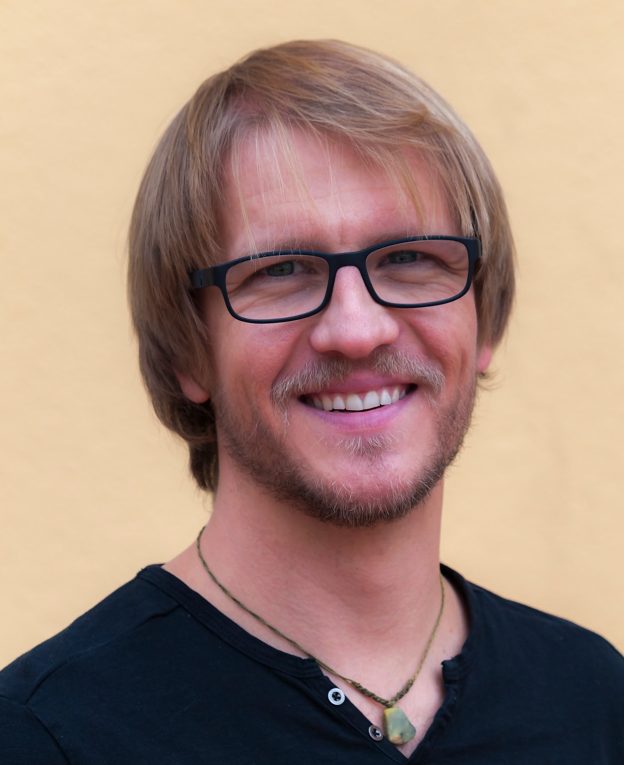}}]{Pau Closas}(Senior Member, IEEE),
is an Associate Professor in Electrical and Computer Engineering at Northeastern University, Boston MA.
He received the M.S. and Ph.D. in Electrical Engineering from UPC in 2003 and 2009, respectively. He also holds a M.S. in Advanced Maths and Mathematical Engineering from UPC since 2014. He is the recipient of the EURASIP Best PhD Thesis Award 2014, the $9^{th}$ Duran Farell Award for Technology Research, the $2016$ ION's Early Achievements Award, $2019$ NSF CAREER Award, and the IEEE AESS Harry Rowe Mimno Award in $2022$. His primary areas of interest include statistical signal processing, stochastic filtering, robust filtering, and machine learning, with applications to positioning and localization systems. He volunteered in editorial roles (e.g. NAVIGATION, Proc. IEEE, IEEE Trans. Veh. Tech., and IEEE Sig. Process. Mag.), and has been actively involved in organizing committees of a number of conference such as EUSIPCO (2011, 2019-2022), IEEE SSP'16, IEEE/ION PLANS (2020, 2023), or IEEE ICASSP'20.
\end{IEEEbiography}

\end{document}